\documentclass[prb,twocolumn,showpacs,superscriptaddress,citeautoscript]{revtex4-1}
\usepackage[latin1]{inputenc}
\usepackage{graphicx}
\usepackage{amssymb}
\usepackage{amsmath}
\usepackage{xspace}
\usepackage{dcolumn}
\usepackage{bm}
\usepackage{color}
\usepackage[extra]{tipa}

\setcitestyle{square,numbers}

\newcommand{\las}[0]{\langle}
\newcommand{\ras}[0]{\rangle}
\newcommand{\llas}[0]{\langle\langle}
\newcommand{\rras}[0]{\rangle\rangle}

\newcommand{\lso}[0]{\lambda}

\newcommand{\rmi}{\text{i}}

\newcommand{\nag}{{\phantom{\dag}}}

\newcommand{\Sxy}{S_\text{AF}^{xy}}
\newcommand{\Sxyz}{S_\text{AF}}
\newcommand{\Uc}{U_\text{c}}

\newfont{\tensy}{cmsy10}
\newcommand{\chem}[1]{{$\fontdimen16\tensy=3.0pt
    \fontdimen17\tensy=3.0pt \mathrm{#1}$}}

\begin{document}

\title{Phase diagram of the Kane-Mele-Coulomb model}

\author{M. Hohenadler}

\affiliation{\mbox{Institut f\"ur Theoretische Physik und Astrophysik,
    Universit\"at W\"urzburg, Am Hubland, 97074 W\"urzburg, Germany}}

\author{F. Parisen Toldin}

\affiliation{\mbox{Institut f\"ur Theoretische Physik und Astrophysik,
    Universit\"at W\"urzburg, Am Hubland, 97074 W\"urzburg, Germany}}

\author{I. F. Herbut}

\affiliation{Department of Physics, Simon Fraser University, Burnaby, British
  Columbia, Canada V5A 1S6}

\author{F. F. Assaad}

\affiliation{\mbox{Institut f\"ur Theoretische Physik und Astrophysik,
    Universit\"at W\"urzburg, Am Hubland, 97074 W\"urzburg, Germany}}

\begin{abstract}
  We determine the phase diagram of the Kane-Mele model with a long-range Coulomb
  interaction using an exact quantum Monte Carlo method.
  Long-range interactions are expected to play a role in honeycomb materials
  because the vanishing density of states in the semimetallic weak-coupling
  phase suppresses screening. According to our results, the
  Kane-Mele-Coulomb model supports the same phases as the Kane-Mele-Hubbard
  model. The nonlocal part of the interaction promotes short-range sublattice charge
  fluctuations, which compete with antiferromagnetic order
  driven by the onsite repulsion. Consequently, the critical interaction for
  the magnetic transition is significantly larger than for the purely local
  Hubbard repulsion. Our numerical data are consistent with $SU(2)$ Gross-Neveu
  universality for the semimetal to antiferromagnet transition, and with 3D XY
  universality for the quantum spin Hall to antiferromagnet transition. 
\end{abstract}

\date{\today}


\maketitle

\section{Introduction}\label{sec:intro}

Inspired by the experimental realization of graphene \cite{Novoselov05},
electrons with a linear band dispersion, or Dirac fermions,
have become a major topic in condensed matter physics. Interest in correlated
fermions on the honeycomb lattice has been boosted by the theoretical
proposal of the quantum spin Hall (QSH) state \cite{KaMe05b}, debates about
the existence of a topological Mott insulator
\cite{RaQiHo08,PhysRevB.89.035103,PhysRevB.88.245123,PhysRevB.89.165123} and a
quantum
spin liquid phase \cite{Meng10,Chen2012,So.Ot.Yu.,Assaad13,Clark2013},
and the mean-field prediction of an interaction-generated topological phase
(the QSH* phase) in a model for \chem{Na_2IrO_3} \cite{Irridates-Nagaosa,PhysRevLett.108.046401}.

In order to make analytical and numerical studies feasible, previous work has
often invoked the approximation of a purely local (Hubbard) repulsion between
electrons \cite{Hu63}. The honeycomb Hubbard model can be simulated
using exact quantum Monte Carlo methods \cite{Sorella92,PhysRevB.72.085123,Meng10}, and has
received considerable interest after reports of a gapped spin liquid phase at
intermediate interactions \cite{Meng10}. Simulations can also  be carried out
for the Hubbard model with additional spin-orbit coupling
\cite{Hohenadler10,Zh.Wu.Zh.11}, usually referred to as the
Kane-Mele-Hubbard (KMH) model \cite{RaHu10}, which provides a framework to
study correlated topological insulators in two dimensions \cite{HoAsreview2013}.

The existence of Dirac cones at isolated points in the Brillouin zone, as
compared to a Fermi surface, is a key feature of the honeycomb lattice
\cite{Neto_rev}. In the absence of interactions and for a half-filled band,
the system is a semimetal (SM) with the density of states vanishing at the
Fermi level \cite{Wallace47}. The SM is stable at weak coupling
\cite{PhysRevLett.97.146401}, and any phase transitions take place at finite
critical interactions. According to analytical calculations, the universality
of the Mott-Hubbard transition should be modified by the presence of
gapless fermionic modes \cite{PhysRevLett.97.146401,Herbut09a,PhysRevB.89.205403,Assaad13}. The
vanishing of the density of states also implies that the Coulomb interaction will
not be screened, and that the approximation of a Hubbard interaction is
therefore {\it a priori} not justified. The long-range Coulomb
interaction leads to a logarithmic divergence of the Fermi velocity
\cite{Gonzalez1994595,PhysRevB.59.R2474} which was confirmed experimentally
\cite{Elias11}, and marginal Fermi liquid behavior
\cite{PhysRevB.59.R2474}. On the other hand, the divergence of the velocity
makes the long-range interaction in graphene marginally irrelevant at the
critical point in the framework of the $\epsilon$-expansion
\cite{Herbut09a}. This result for weak interactions, which holds close to
$3+1$ dimensions \cite{PhysRevB.59.R2474,PhysRevLett.97.146401,Herbut09a} as well
as close to $1+1$ dimension \cite{PhysRevB.80.081405}, suggests the same universality class for the
Mott transition as in the Hubbard model.

For the Hubbard model, our present understanding based on numerical and
analytical results suggests the existence of a second-order Mott transition from a semimetallic to
an antiferromagnetic phase at $U/t\approx3.8$, with no intermediate spin
liquid phase \cite{So.Ot.Yu.,Assaad13}. The KMH model with
additional spin-orbit coupling instead undergoes a transition from a quantum
spin Hall (QSH) state to an antiferromagnetic phase, also at a finite critical
$U$ \cite{RaHu10,Hohenadler10,Zh.Wu.Zh.11}. The phase diagram is shown in Fig.~\ref{fig:phasediagram}(b). Quantum
Monte Carlo (QMC) data are consistent with the predicted $SU(2)$ Heisenberg Gross-Neveu universality
for the Hubbard model \cite{Assaad13}, and with 3D XY universality for the
KMH model \cite{Ho.Me.La.We.Mu.As.12}. The honeycomb lattice
with long-range Coulomb interaction has been investigated in detail in the
context of graphene, see \cite{Neto_rev} for a review. An interaction-driven
metal-insulator transition in graphene was demonstrated using quantum Monte Carlo simulations
\cite{PhysRevLett.102.026802}.  More recently, long-range Coulomb interaction
has been studied in models with Dirac and Weyl fermions
\cite{PhysRevB.87.165142,PhysRevB.87.205440,Sekine13,Sekine14}, including the
KM model \cite{PhysRevB.87.205440}.

In this work, we present exact results for electrons on the honeycomb lattice
interacting via a $1/r$ Coulomb potential. The auxiliary-field QMC method
used is free of a sign problem at half filling, and can be
applied to more general nonlocal interactions. Here, we study the phase
diagram of the Kane-Mele-Coulomb model. In the absence of spin-orbit
coupling, we find a quantum phase transition from an SM to
an antiferromagnet consistent with the Gross-Neveu universality class, with
the critical point shifted to larger interaction strengths compared to the
Hubbard model. At a nonzero spin-orbit coupling, the KM model with
long-range electron-electron interaction is found to be either in a QSH
or in a magnetic insulating state. Similar to the KMH model, the
phase transition is consistent with the 3D XY universality class, but again
occurs at larger values of the interaction. We find no evidence of a potential,
intermediate QSH* phase \cite{PhysRevLett.108.046401} or any other additional phases.

The paper is organized as follows. In Sec.~\ref{sec:model}, we define the
models. Section~\ref{sec:method} provides a discussion of the QMC
method. Our results are presented in Sec.~\ref{sec:results}, and
Sec.~\ref{sec:conclusions} contains our conclusions.

\section{Kane-Mele-Coulomb model}\label{sec:model}

The KM Hamiltonian \cite{KaMe05a,KaMe05b} can be written as
\begin{equation}\label{eq:KM}
  \hat{H}_{0} 
  = 
  -t \sum_{\las \bm{i},\bm{j} \ras} \hat{c}^{\dagger}_{\bm{i}} \hat{c}^\nag_{\bm{j}}
  + 
  \rmi\,\lso \sum_{\llas \bm{i},\bm{j}\rras}
  \hat{c}^{\dagger}_{\bm{i}}\,
  (\boldsymbol{\nu}_{\bm{i}\bm{j}} \cdot \boldsymbol{\sigma})\,
  \hat{c}^\nag_{\bm{j}} \,.
\end{equation}
Here, we have used the spinor notation ${\hat{c}^{\dagger}_{\bm{i}} = \big(c^{\dagger}_{\bm{i}\uparrow},
  c^{\dagger}_{\bm{i}\downarrow}\big)}$, where $c^{\dagger}_{\bm{i}\sigma}$ creates an
electron with spin $\sigma$ at site $\bm{i}$. The symbols $\las
\bm{i},\bm{j}\ras$ and $\llas \bm{i},\bm{j}\rras$ denote pairs of nearest-neighbor and
next-nearest-neighbor lattice sites on the honeycomb lattice, respectively,
and implicitly include the Hermitian conjugate terms. The first term
corresponds to the usual nearest-neighbor hopping \cite{Neto_rev}. The second
term describes the $z$ component of Rashba spin-orbit coupling
\cite{KaMe05a} in graphene, which takes the form of a complex next-nearest
neighbor hopping $\pm \rmi\lambda$. The sign depends on the sublattice, the
electron spin, and the direction of the hopping process. It may be compactly
written in the form $\boldsymbol{\nu}_{\bm{i}\bm{j}} \cdot \boldsymbol{\sigma}$, with
\begin{equation}
  \boldsymbol{\nu}_{\bm{i}\bm{j}} = \frac{\bm{d}_{\bm{i}\bm{k}} \times
    \bm{d}_{\bm{k}\bm{j}}}{|\bm{d}_{\bm{i}\bm{k}} \times \bm{d}_{\bm{k}\bm{j}}|}\,.
\end{equation}
The vector $\bm{d}_{\bm{i}\bm{k}}$ (with vanishing $z$ component) connects sites
$\bm{i}$ and $\bm{k}$, $\bm{k}$ being the intermediate lattice site between
$\bm{i}$ and $\bm{j}$; $\boldsymbol{\sigma} = (\sigma^x,\sigma^y,\sigma^z)$
is the Pauli vector.

The QMC method used here can be applied to a rather general electron-electron
interaction of the form
\begin{equation} \label{LRI.eq}
  \hat{H}_V =  \frac{1}{4} \sum_{\bm{i}\bm{j}}
  V_{\bm{i}\bm{j}}  ( \hat{n}_{\bm{i}}   - 1 )  
  (  \hat{n}_{\bm{j}}   - 1) \,,
\end{equation}
with a positive-definite matrix $V$.
The numerical results shown were obtained for the specific choice
\begin{equation}\label{eq:Vused}
  V_{\bm{i}\bm{j}}   =   \left\{
    \begin{array}{cc}
      2 U\,, & \text{if  } | \bm{i} - \bm{j} | = 0 \\
      \frac{\alpha U\delta}{ | \bm{i} - \bm{j} | }  \,,  &  \text{if  } | \bm{i}
      - \bm{j} | >0
    \end{array}
  \right.\,.
\end{equation}
In Eq.~(\ref{eq:Vused}), $\alpha$ determines the relative strength of the 
onsite and the nonlocal interactions,
and $\delta = \frac{2}{3}| \bm{a}_2 - \frac{1}{2}\bm{a}_1| $ is
the distance between the two orbitals in the unit cell [$\bm{a}_1 =
(1,0) $, $\bm{a}_2 = \frac{1}{2}\left( 1, \sqrt{3} \right) $ are the basis
vectors of the honeycomb lattice].  The distance $ | \bm{i} - \bm{j} | $ is
the minimal distance between the sites $\bm{i} $ and $\bm{j}$.
For $\alpha=0$, $\hat{H}_V$ reduces to the Hubbard interaction
\begin{equation}
  \hat{H}_U = \frac{U}{2} \sum_{\bm{i}}\left(\hat{n}_{\bm{i}} -1 \right)^2\,.
\end{equation}

We refer to the Hamiltonian $\hat{H}=\hat{H}_0+\hat{H}_V$, with
$V_{\bm{i}\bm{j}}$  defined as in Eq.~(\ref{eq:Vused}) as the {\it
  Kane-Mele-Coulomb} (KMC) model. Its Hamiltonian respects $C_3$ rotational
symmetry, $U(1)$ spin symmetry, $Z_2$ time-reversal
symmetry, and $U(1)$ gauge invariance. In the absence of spin-orbit coupling
($\lso=0$), we recover the full $C_6$ rotation symmetry of the lattice, and
$SU(2)$ spin symmetry. At half filling, there is an additional particle-hole symmetry.

Throughout the paper we will consider half-filled lattices with $L\times L$ unit cells
and periodic boundary conditions. The number of lattice sites is given by $N=2L^2$.

\section{Quantum Monte Carlo Method}\label{sec:method}

We discuss the method for a Hamiltonian $\hat{H}=\hat{H}_0+\hat{H}_V$, with $\hat{H}_0$
given by Eq.~(\ref{eq:KM}) and a general, nonlocal interaction as defined by Eq.~(\ref{LRI.eq}).

The starting point for the implementation of the long-range interaction is the action
\begin{eqnarray}
  S(\left\{ A, c^{\dag}, c \right\})  
  &=&  S_0(\left\{c^{\dag}, c \right\})  \\\nonumber
  && + \int_{0}^{\beta} {\rm d} \tau \sum_{\bm{i}}   \rmi A(\bm{i},\tau) \left[ n_{\bm{i}\sigma}(\tau)    - 1 \right]  \\\nonumber
  && +   \int_{0}^{\beta} {\rm d} \tau  \sum_{\bm{i}\bm{j}} A(\bm{i},\tau)  V^{-1}_{\bm{i}\bm{j}} A(\bm{j},\tau)\,.
\end{eqnarray}
Here, $S_0$ corresponds to the action of the noninteracting
Hamiltonian~(\ref{eq:KM}), $n_{\bm{i}} (\tau)= \sum_{\sigma}
c^{\dag}_{\bm{i}\sigma}(\tau) c^{\nag}_{\bm{i}\sigma}(\tau)$, and
$A(\bm{i},\tau) $ is a real scalar field.  If the matrix
$V_{\bm{i}\bm{j}}$ is positive definite, the Gaussian integral over
the scalar field can be carried out to give 
\begin{eqnarray}
  S(\left\{c^{\dag}, c \right\})
  &=&
  S_0(\left\{c^{\dag}, c \right\}) + S_1(\left\{c^{\dag}, c \right\})  
\end{eqnarray}
with 
\begin{eqnarray}  
  S_1(\left\{c^{\dag}, c \right\})  
  &=&\frac{1}{4}\! \int_{0}^{\beta} {\rm d} \tau  \sum_{\bm{i}\bm{j}}   [ n_{\bm{i}}(\tau)    - 1 ]   V_{\bm{i}\bm{j}} [  n_{\bm{j}}(\tau)    - 1]\,.
\end{eqnarray}
A similar approach was used in Ref.~\cite{Ulybyshev2013}.

The action in the presence of the scalar field is quadratic in the fermionic
degrees of freedom.  The latter can hence be integrated out to obtain
\begin{eqnarray}
  S(\left\{ A\right\})   =  & &  \int_{0}^{\beta} {\rm d} \tau  \sum_{\bm{i}\bm{j}} A(\bm{i},\tau)  V^{-1}_{\bm{i}\bm{j}} A(\bm{j},\tau)      \nonumber \\
  & &  -\ln \text{Tr} \left[ {\cal T} e^{- \int_{0}^{\beta} \text{d} \tau  \hat{H}(\left\{ A\right\})} \right]\,
\end{eqnarray}  
where
\begin{equation}
  \hat{H}(\left\{ A\right\})  = \hat{H}_0  + \sum_{\bm{i}}   \rmi A(\bm{i},\tau) \left( \hat{n}_{\bm{i}}    - 1 \right)\,.
\end{equation}

The presence of particle-hole and $U(1)$ spin symmetry guarantees that the action is real.
In particular, the $U(1)$ spin symmetry allows us to factorize the trace into
spin-up and spin-down contributions,
\begin{equation}
  \text{Tr} \left[ {\cal T} e^{- \int_{0}^{\beta} \text{d} \tau  \hat{H}(\left\{ A\right\})} \right] = 
  \prod_{\sigma} \text{Tr}_{\sigma} \left[ {\cal T} e^{- \int_{0}^{\beta} \text{d} \tau  \hat{H}_{\sigma}(\left\{ A\right\})} \right] \,.
\end{equation}
With the canonical transformation $c^{\dag}_{\bm{i}\uparrow} \rightarrow
(-)^{\bm{i}} c^{\nag}_{\bm{i}\downarrow}$, where $(-)^{\bm{i}}$ takes the
value $1$ ($-1$) on the $A$ ($B$) sublattice, we can show that (the bar
denotes complex conjugation)
\begin{equation}
  \text{Tr}_{\uparrow} \left[ {\cal T} e^{- \int_{0}^{\beta} \text{d} \tau  \hat{H}_{\uparrow}(\left\{ A\right\})} \right]  =
  \overline{\text{Tr}_{\downarrow} \left[ {\cal T} e^{- \int_{0}^{\beta} \text{d} \tau  \hat{H}_{\downarrow}(\left\{ A\right\})} \right] }\,.
\end{equation}
Therefore, the action $S(\left\{ A\right\}) $ is real and the weight for a
given field configuration, $ e^{-S(\left\{ A\right\}) }$, is positive.
Consequently, the Monte Carlo sampling of the scalar field does not suffer
from the minus sign problem.

The implementation of the method relies on a Trotter discretization of
imaginary time, $\beta = L_\tau \Delta \tau$.  There are many possibilities
for carrying out the sampling. A possible choice is hybrid
molecular dynamics \cite{Ulybyshev2013,PhysRevLett.102.026802,Scalettar87}
based on a Gaussian integral representation of the determinant.  Here, we
have implemented a  simpler, sequential updating scheme in which field
configurations are proposed according to the probability $
e^{\int_{0}^{\beta} {\rm d} \tau  \sum_{\bm{i},\bm{j}} A(\bm{i},\tau) V^{-1}_{\bm{i}\bm{j}}
  A(\bm{j},\tau) } $ and then accepted or rejected using importance sampling.
This approach is advantageous when the matrix $V$ has a small number of
low-lying eigenvalues that favor specific modulations of the scalar field.
To implement the algorithm, we chose a basis where $V$ is diagonal. Because $V$ is symmetric and
positive definite, we can find an orthogonal transformation $M$ such that
$M^{\dag} V M= \text{diag}\left( \xi_1, \cdots, \xi_N\right)$ with
$\xi_{\bm{i}} > 0 $.  With $ \Phi(\bm{i},\tau) = \sum_{\bm{j}}
M^{\dag}_{\bm{i} \bm{j} } A(\bm{j},\tau) $, the partition function reads as
\begin{equation}
  Z   
  = \int {\text D} \left\{ \Phi \right\}  \prod_{\bm{i}, \tau}   e^{
    - \Delta \tau {\Phi^2(\bm{i},\tau) } /{\xi_{\bm{i}}}   }  
  W\left( \left\{ \Phi \right\} \right)    
  + { \cal O}\left( \Delta \tau ^2 \right)
\end{equation}
where
\begin{eqnarray}
  W\left( \left\{ \Phi \right\} \right)
  &=&    
  \text{Tr}
  \prod_{\tau}  e^{-\frac{\Delta\tau}{2}\hat{H}_0 } 
  \\\nonumber
  & &\quad\quad\times  e^{- \Delta \tau \sum_{\bm{i}\bm{j}}   \rmi
    M_{\bm{i}\bm{j}} \Phi(\bm{j},\tau) \left( \hat{n}_{\bm{i}}
      - 1 \right) }   e^{-\frac{\Delta \tau}{2}\hat{H}_0}\,.
\end{eqnarray}

We propose new configurations according to
\begin{eqnarray}\label{eq:T0}
  T_0\left( \left\{ \Phi \right\}  \rightarrow \left\{ \Phi' \right\}\right)  
  &=&  \prod_{\bm{i},\tau} \big\{  P_{\bm{i}\tau}    P_0[
    \Phi'(\bm{i},\tau)]     \\\nonumber
  & &\,\,\,+   
    (1-P_{\bm{i}\tau} )   \delta  [  \Phi_{0}( \bm{i},\tau) -    \Phi'_{0}( \bm{i},\tau)  ]  
    \big\}\,,
\end{eqnarray}
with $P_0( \Phi'(\bm{i},\tau) ) = \sqrt{\frac{\Delta \tau}{\pi
    \xi_{\bm{i}}}} e^{-\Delta \tau \Phi'^2(\bm{i},\tau)
  /\xi_{\bm{i}} } $. The proposed configuration $\left\{ \Phi' \right\}$ is
accepted with probability
\begin{eqnarray}\nonumber
  P &=& 
  \text{min} \left(\frac{T_0\left( \left\{ \Phi' \right\}  \rightarrow \left\{ \Phi \right\} \right)  W_\text{tot} (\left\{ \Phi' \right\} )  }
    { T_0\left( \left\{ \Phi \right\}  \rightarrow \left\{ \Phi' \right\} \right)  W_\text{tot} (\left\{ \Phi \right\} ) } , 1 \right)  \\
  &\equiv& \text{min}  \left( \frac{W (\left\{ \Phi' \right\} ) }{W (\left\{ \Phi \right\} ) } , 1 \right). 
\end{eqnarray} 
Here, the total weight of a configuration is given by
\begin{equation}
  W_\text{tot} ( \left\{ \Phi \right\} )  = \prod_{\bm{i},\tau} P_0\left( \Phi(\bm{i},\tau)\right) W (\left\{ \Phi \right\} )\,.
\end{equation}
The probabilities $P_{\bm{i},\tau}$ in Eq.~(\ref{eq:T0}) can be chosen arbitrarily, allowing us to
optimize the acceptance rate. We have opted for a sequential updating of the
time slices. We set $ P_{\bm{i},\tau} \equiv P $ and used values of $P$ that
yield a good acceptance rate for updates.

Because we are interested in ground-state properties, we used the
projective (zero-temperature) auxiliary-field QMC
algorithm.  Taking $ | \Psi_T \rangle $ to be the ground state of the
noninteracting Hamiltonian $\hat{H}_0$, and assuming that it has a finite
overlap with the ground state $| \Psi_0 \rangle $ of $\hat{H}$,
expectation values can be calculated as
\begin{equation}
  \frac{ \langle  \Psi_0 | \hat{O} |  \Psi_0 \rangle  }{ \langle  \Psi_0  |  \Psi_0 \rangle  }  =  
  \lim_{\Theta \rightarrow \infty} \frac{ \langle  \Psi_\text{T} | e^{-\Theta \hat{H} /2 } \hat{O} e^{-\Theta \hat{H} /2  } |  \Psi_\text{T}\rangle  }
  { \langle  \Psi_\text{T}  | e^{-\Theta \hat{H} }  |  \Psi_\text{T} \rangle  } .
\end{equation}
The implementation of the projective algorithm is similar to that for
finite temperatures, a detailed description of which can be found in
Ref.~\cite{Assaad08_rev}. Dynamical correlation
functions were computed with the method of Ref.~\cite{Feldbach00}.  We used a symmetric Trotter
decomposition to minimize the systematic error, with $\Delta \tau t = 0.1$. A projection parameter
$\Theta t = 40$ was sufficient to achieve convergence to the ground state
within statistical errors.  

\begin{figure}[t]
  \includegraphics[width=0.5\textwidth]{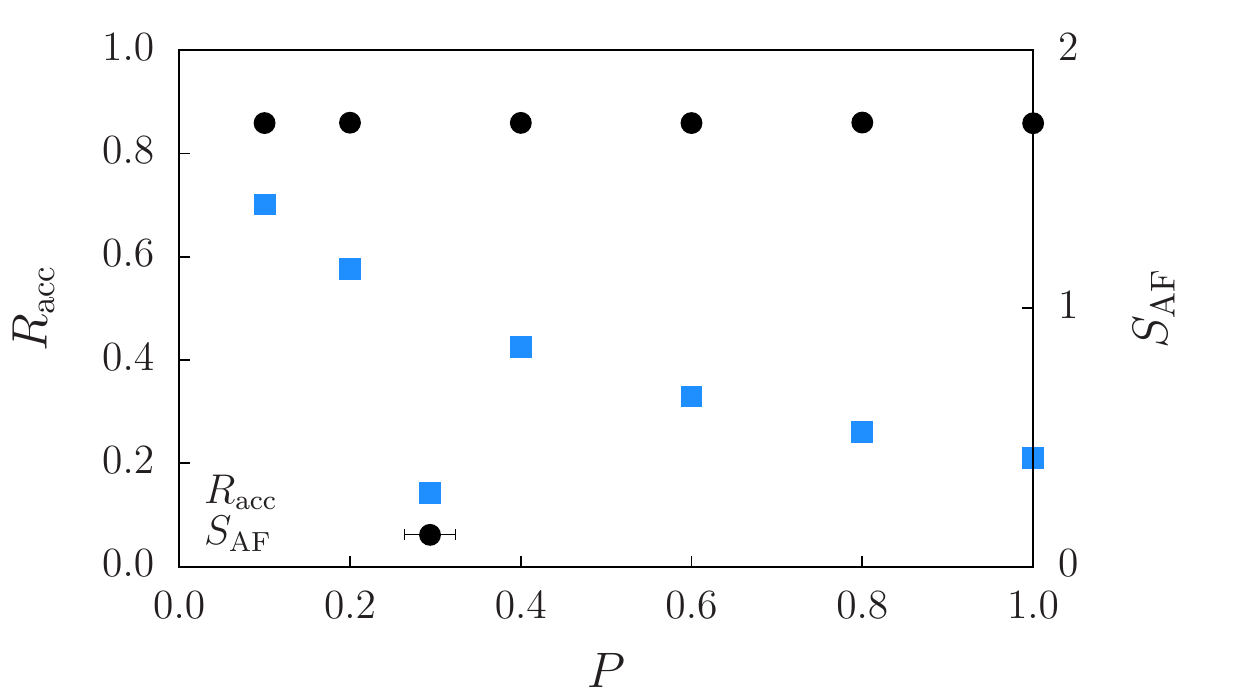}
  \caption{\label{fig:Test} (Color online) Acceptance rate $R_\text{acc}$ for
    Monte Carlo updates and spin structure factor $S_\text{AF}$ [Eq.~(\ref{eq:S})] as a function
    of the parameter $P$. Here, $U/t=4$, $\alpha=1$, $L=6$.}
\end{figure}

Figure~\ref{fig:Test} shows the acceptance rate $R_\text{acc}$
and the spin structure factor [Eq.~(\ref{eq:SAF})] as a function of the
parameter $P$ for $U/t=4$ and $\alpha=1$. The results reveal that $P$ can be used to tune the acceptance rate without changing the
values of physical observables. 

Compared to the Hubbard interaction, simulations
with the long-range interaction~(\ref{eq:Vused}) and local updates become
increasingly difficult at strong interactions, leading to long
autocorrelation times. Because the phase transitions in the KMC model occur at
larger interactions, the quality of the data and the finite-size
extrapolations is not as good as for the KMH model
\cite{Meng10,Assaad13,Hohenadler10,Ho.Me.La.We.Mu.As.12}.

\section{Results}\label{sec:results}

To better orient the discussion, we first present the phase diagram of the
KMC model. Then, we discuss how the phase boundaries were
obtained from finite-size scaling, look at the critical behavior, provide an
explanation for the shift of the magnetic transition compared to the
KMH model, and comment on the absence of new phases. We focus on
$\alpha = 1$, but very similar results were obtained for $\alpha=1.23$.

\subsection{Phase diagram}\label{sec:phasediagram}

\begin{figure}[t]
  \includegraphics[width=0.45\textwidth]{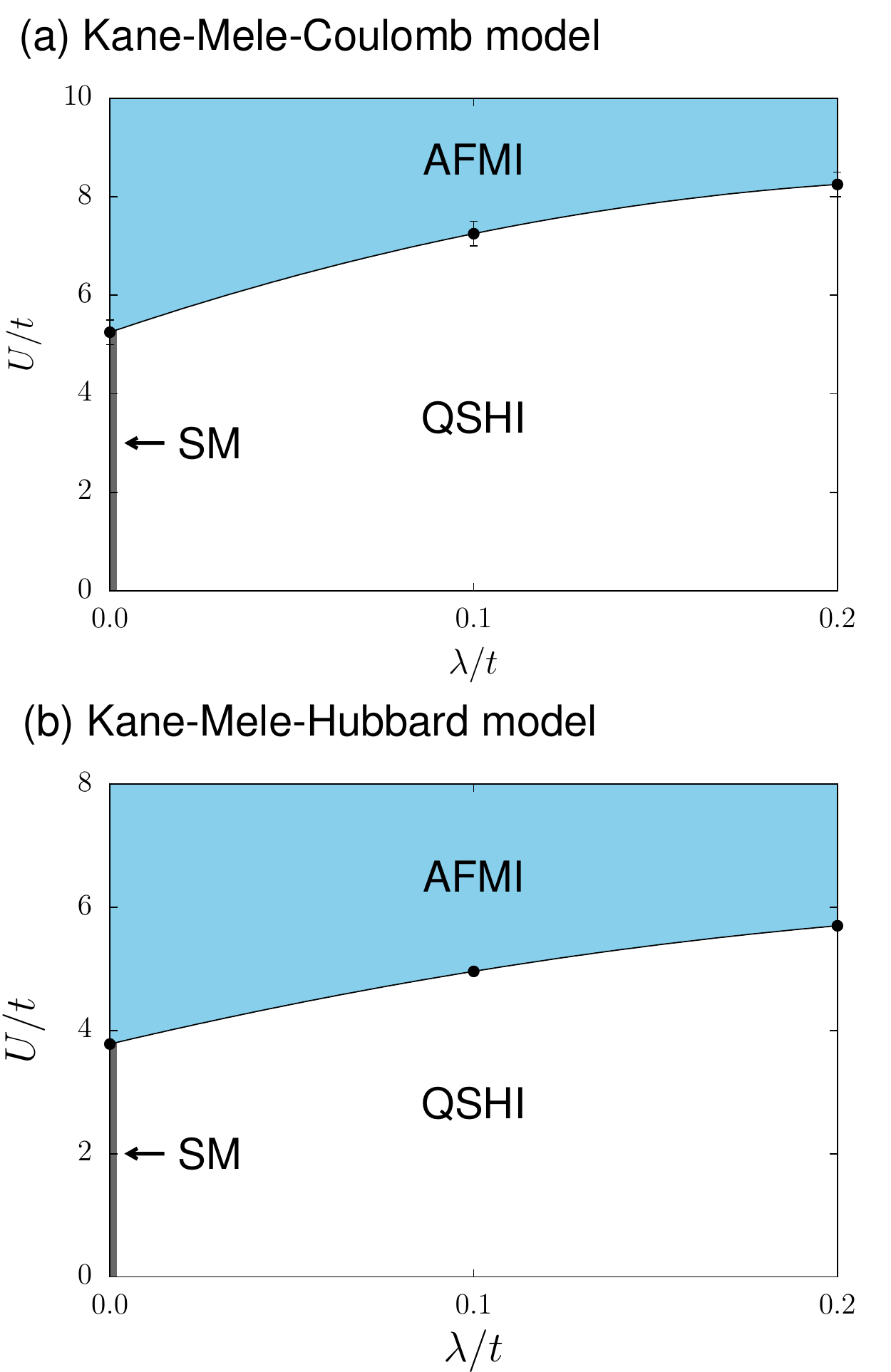}
  \caption{\label{fig:phasediagram} (Color online) (a) Phase diagram of the
    KMC model with $\alpha=1$. The phases correspond to a semimetal (SM)
    which exists for $\lambda=0$, a quantum spin Hall insulator (QSHI), and
    an antiferromagnetic Mott insulator (AFMI). 
    (b) Phase diagram of the KMH model (corresponding to $\alpha = 0$) based on previous
    simulations \cite{Hohenadler10,Ho.Me.La.We.Mu.As.12,As.Be.Ho.2012,Assaad13},
    see also Ref.~\cite{inside2013}.}
\end{figure}

The zero-temperature phase diagram of the KMC model with $\alpha=1$, as
obtained from QMC simulations, is shown in Fig.~\ref{fig:phasediagram}(a).  For
comparison, we also show the phase diagram of the KMH model in
Fig.~\ref{fig:phasediagram}(b). As discussed in detail below, the phase boundaries are
based on a finite-size scaling of the magnetization $m$. 
The data for the KMH model were taken from
Refs.~\cite{Hohenadler10,Ho.Me.La.We.Mu.As.12,As.Be.Ho.2012,Assaad13}.
The restriction of the SM to $\lambda=0$ follows from the fact
that the spin-orbit term immediately opens a mass gap in a gapless Dirac
metal, as previously illustrated for the KMH model \cite{Ho.Me.La.We.Mu.As.12}.

Similar to the KMH model \cite{Meng10}, the KMC model has a semimetallic
ground state for $\lambda=0$ and $U<U_\text{c}$; note that with our definition of the
interaction in Eq.~(\ref{eq:Vused}), both the local and the nonlocal part of
the interaction scale with $U$. For stronger interactions $U>U_\text{c}$, the
ground state is an antiferromagnetic Mott insulator (MI). At nonzero spin-orbit
coupling $\lambda$, we find a QSH phase up to a critical
$U_\text{c}$, and again a magnetic insulator for $U>U_\text{c}$. The same
phases have previously been observed in the KMH model
\cite{RaHu10,Hohenadler10,Zh.Wu.Zh.11}, see Fig.~\ref{fig:phasediagram}(b).
As for the KMH model, the critical value for the transition increases with
increasing $\lambda$. We have found no evidence for a previously reported
quantum spin liquid phase at intermediate
interactions \cite{Meng10,Hohenadler10,Ho.Me.La.We.Mu.As.12}, consistent with
recent numerical results for the Hubbard
model \cite{So.Ot.Yu.,Assaad13,Clark2013}. 

As observed before for the KMH model, the magnetic ordering in the AFMI
occurs in the transverse and longitudinal spin directions at $\lambda=0$, whereas only the
transverse spin components order at $\lambda>0$. For the KMH model, the
effective spin model \cite{RaHu10} valid at large $U/t$ contains exchange
interactions $J=4t^2/U$ and $J'=\pm4\lambda^2/U$. The sign of $J'$ is different
for the $z$ ($J'>0$) and the $xy$ ($J'<0$) directions of spin. Because $J$
and $J'$ act between nearest-- and next-nearest-neighbor spins, respectively,
the $z$ component becomes frustrated for $\lambda\neq 0$, favoring easy-plane
antiferromagnetic order. The different symmetry of the order parameter at $\lambda=0$ and
$\lambda>0$, and the absence or presence of gapless fermionic modes below
$U_\text{c}$, also implies different universality classes for the corresponding
phase transitions. Numerical results for the KMH model are consistent with an
$SU(2)$ Gross-Neveu transition for $\lambda=0$ \cite{Assaad13}, and a $U(1)$
3D XY transition for $\lambda>0$ \cite{Ho.Me.La.We.Mu.As.12}.

\subsection{Magnetic phase transition at $\lambda=0$}\label{sec:SM-Mott}

\begin{figure}[t]
  \includegraphics[width=0.45\textwidth]{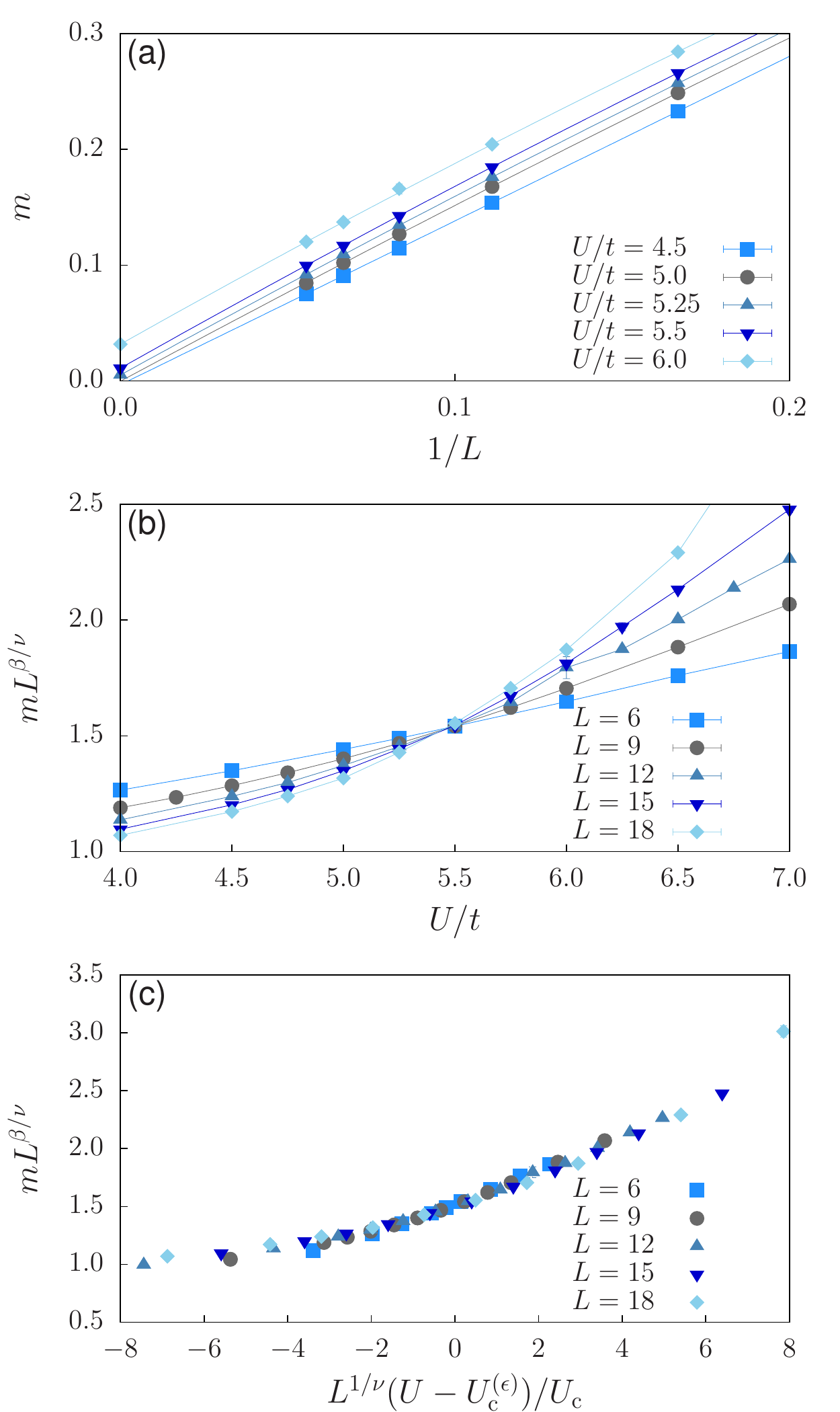}  
  \caption{\label{fig:Collapse_V1} (Color online)
    SM--AFMI quantum phase transition.  (a)
    Finite-size scaling of the magnetization $m$ using quadratic fits. (b) Scaling
    intersection using the critical exponents for the Gross-Neveu
    universality class from the $\epsilon$ expansion
    \cite{Herbut09a,Assaad13}.  (c) Scaling collapse using the critical value
    $\Uc^{(\epsilon)}/t=5.45$. Here, $\lambda=0$, $\alpha=1$.
   }
\end{figure}

\begin{figure}[t]
  \includegraphics[width=0.45\textwidth]{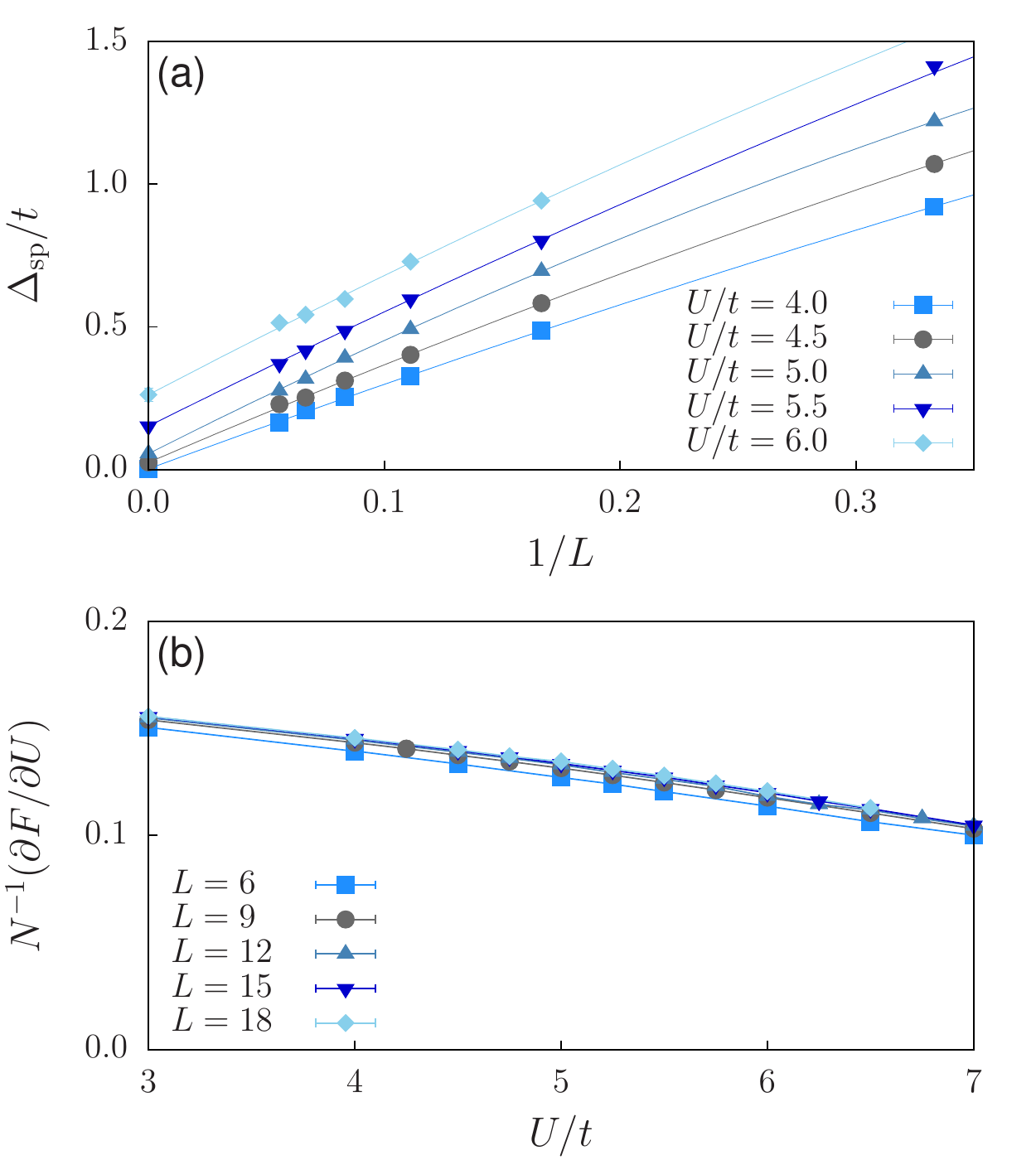}  
  \caption{\label{fig:spgap_lambda0.0} (Color online)
    (a) Finite-size scaling of the single-particle gap using quadratic
    fits. (b) Expectation value of the interaction term, corresponding to the
    derivative of the free energy with respect to $U$.  Here,
    $\lambda=0$, $\alpha=1$.}
\end{figure}

The quantum phase transition from the SM to the AFMI in the Hubbard model on the honeycomb lattice
($\alpha=0$, $\lambda=0$) has attracted a lot of interest, partly
because the transition has a finite critical value $\Uc$, and can be studied
exactly using QMC methods. An intriguing question is if
the transition between these phases is a direct transition
\cite{So.Ot.Yu.,Assaad13,Clark2013} or involves an intermediate spin liquid
phase \cite{Meng10,Chen2012}. After initial evidence for the existence of
such a phase  \cite{Meng10}, more recent results on larger lattices
\cite{So.Ot.Yu.} and using alternative methods to compute the order
parameter \cite{Assaad13} favor the scenario of a direct quantum
phase transition. The absence of a jump in the double occupation (which
corresponds to the derivative of the free energy with respect to $U$) at the
critical point suggests that the transition is continuous \cite{Meng10}.

In Ref.~\cite{Assaad13}, it was shown that QMC results for the Mott
transition of the honeycomb Hubbard model are consistent
with a novel fermionic critical point described by the Gross-Neveu-Yukawa theory
\cite{Herbut09,Herbut09a}.  The latter describes Dirac fermions coupled to
magnetic (bosonic) degrees of freedom via a Yukawa term \cite{Herbut09a}.
The question we address here is if the nature of the transition is altered by
a long-ranged Coulomb interaction.  Analytically, it is possible to
include the Coulomb potential with the help of a scalar field and show that
it is, if weak, a (marginally) irrelevant perturbation \cite{Herbut09a}.  

To study the onset of long-range antiferromagnetic order, we consider the
spin-spin correlation function
\begin{equation}\label{eq:SdotS}
  S_{\alpha\beta}(\bm{i}-\bm{j}) 
  = 
  \las \bm{S}_{\bm{i}} \cdot \bm{S}_{\bm{j}} \ras\,,
\end{equation}
where $\alpha$ ($\beta$) is the orbital index belonging to site $\bm{i}$ ($\bm{j}$),
and the corresponding $\bm{Q}=0$ structure factor 
\begin{equation}\label{eq:S}
  \Sxyz
  = \frac{1}{L^2}
  \sum_\alpha
  \sum_{\bm{i}\,\bm{j}} 
  S_{\alpha\alpha}(\bm{i}-\bm{j})\,,
\end{equation}
where we have taken the trace over the orbitals. The magnetization per
site is then given by
\begin{equation}\label{eq:mxyz}
  m = \sqrt{\Sxyz/N}\,.
\end{equation}
It extrapolates to zero in the nonmagnetic SM phase, but takes on a finite
value in the thermodynamic limit for $U\geq\Uc$, where $\Uc$ is the critical
value for the magnetic phase transition. In contrast to previous work
\cite{Assaad13}, we did not use pinning fields.

Figure~\ref{fig:Collapse_V1}(a) shows the finite-size scaling of the
magnetization for different values of $U/t$. We simulated system sizes
ranging from $L=6$ to $L=18$, and used quadratic fits for the
extrapolation. Within the accuracy of this scaling procedure, the phase
transition seems to occur between $U/t=5$ and $U/t=5.5$. The critical value
is hence significantly larger than for
the transition in the Hubbard model where $\Uc/t=3.78(5)$ \cite{Assaad13}.

As for the Hubbard model \cite{Assaad13}, we test if our data are compatible
with the critical exponents $z=1$, $\beta/\nu=0.9$ and $\nu=1/2 +
21/55\approx0.88$ for the $SU(2)$ Gross-Neveu universality class in 2+1
dimensions, obtained from the $\epsilon$-expansion with $\epsilon=1$.
The plot of $m L^{\beta/\nu}$ in Fig.~\ref{fig:Collapse_V1}(b)
produces a satisfactory intersection of curves for different system sizes at
a critical value $\Uc^{(\epsilon)}/t=5.45(10)$. 

Using $\Uc^{(\epsilon)}/t=5.45$, we plot $m L^{\beta/\nu}$ as a function of
$L^{1/\nu}(U-\Uc)/\Uc$ in Fig.~\ref{fig:Collapse_V1}(c). The rather good
scaling collapse suggests that our numerical data are consistent with the
Gross-Neveu critical exponents from the $\epsilon$-expansion, similar to the
analogous transition in the Hubbard model  \cite{Assaad13}. The scaling
collapse quickly deteriorates upon variation of $\Uc^{(\epsilon)}$.

Figure~\ref{fig:spgap_lambda0.0}(a) shows a finite-size scaling of the
single-particle excitation gap $\Delta_\text{sp}$, which is extracted from
fits to the single-particle Green function at the Dirac point,
$G(\bm{q}=\bm{K},\tau)$ \cite{Meng10,Ho.Me.La.We.Mu.As.12}. Second-order
polynomial extrapolations to the thermodynamic limit suggest a vanishing of
the gap for $U/t\leq 4.5$, and a very small but nonzero single-particle gap
for $U/t\geq 5.5$. For $U/t=5$, the data curve downward at large $L$; a
quadratic fit suggests a small but nonzero gap.

The uncertainty in the finite-size extrapolation of $m$ and
$\Delta_\text{sp}$ is larger than for the KMH model. In particular, these
quantities suggest a critical value in the range $[5,5.5]$, smaller than
$\Uc^{(\epsilon)}/t=5.45(10)$ obtained using the critical exponents from the
$\epsilon$ expansion. Apart from the limitations in system size, which affect
the accuracy of the finite-size extrapolation of $m$ and $\Delta_\text{sp}$,
it was previously shown that a measurement of $m^2$ instead of $m$ is
problematic close to the critical point \cite{Assaad13}. In addition, we see
evidence for logarithmic corrections to scaling for the system sizes
considered. The critical value $\Uc^{(\epsilon)}/t=5.45(10)$ further depends
on the ratio $\beta/\nu$, with $\beta$ and $\nu$ obtained from the $\epsilon$
expansion. The accuracy of the values for the critical exponents is unknown
for the present model, but a recent comparison with QMC simulations for $Z_2$
and $U(1)$ Gross-Neveu models showed good agreement
\cite{PhysRevD.88.021701,PhysRevB.87.041401}. A scaling analysis
independent of critical exponents is beyond the scope of this paper, and will
be published elsewhere.

Finally, Fig.~\ref{fig:spgap_lambda0.0}(b) shows the free-energy
derivative $\partial F/\partial U=\las\hat{H}_V\ras/U$, see also Eq.~(\ref{eq:Vused}). The
continuous evolution of this quantity across the critical point suggests a
continuous (second-order) phase transition.

\subsection{Competition of spin and charge order}\label{sec:comp}

The main difference between the phase diagrams of the KMC and the KMH model
is a shift of the magnetic phase to larger values of $U$. For $\lambda=0$,
this shift can  be understood already at the classical level where the
total energy is given by
\begin{equation}
  E_\text{cl}    = \frac{1}{4}  \sum_{\bm{i}\bm{j}}  ( n_{\bm{i}} -1 ) V_{\bm{i}\bm{j}} ( n_{\bm{j}} -1 )\,.
\end{equation}
The state with uniform density at half filling (that is, with $n_{\bm{i}}
=1$)  has $E_\text{cl}=0$. If $V_{\bm{i}\bm{j}}$ is positive definite, all other charge
configurations have a positive and hence higher energy. However, with increasing $\alpha$, 
the energy of the charge-density-wave state with a doubly occupied sites on
sublattice A and empty sites on sublattice B (or vice versa) decreases,
leading to a competition with the uniform state. For a model with onsite
($U$) and nearest-neighbor ($V$) repulsion only, the two states
become degenerate when $3V=U$, whereas for the long-range
interaction~(\ref{eq:Vused}) degeneracy occurs close to  $\alpha=1.23$.
The competition between the magnetic Mott state and the charge-density-wave
state provides an explanation for the observed increase of the critical value
$\Uc$ upon going from a Hubbard to a long-range interaction, see Fig.~\ref{fig:phasediagram}.
The suppression of magnetic order can also be understood as resulting from a
reduction of the effective onsite repulsion by the nonlocal interactions \cite{Schuler13}.

\begin{figure}[t]
  \begin{center}
    \includegraphics[width=0.5\textwidth]{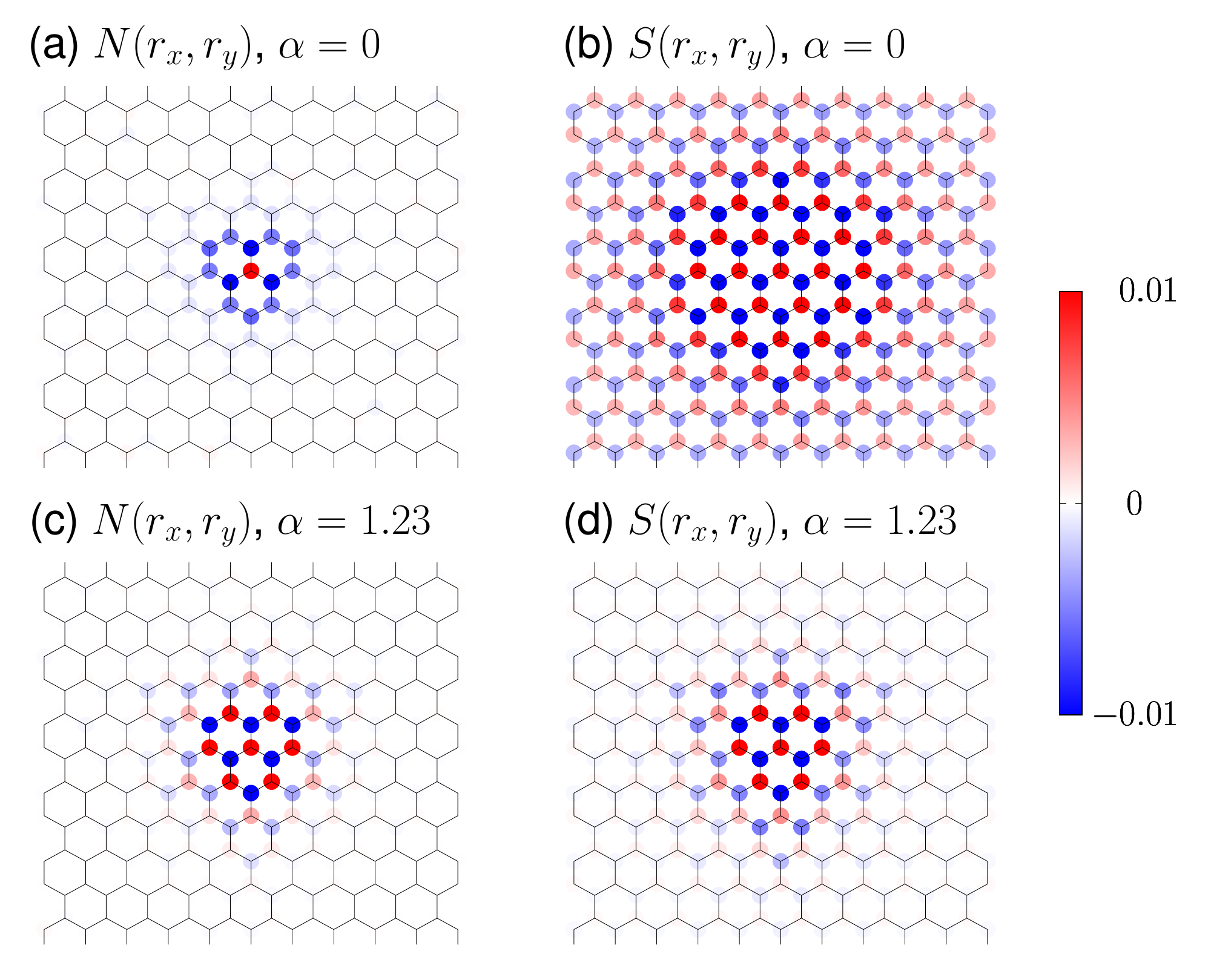}
  \end{center}
  \caption{\label{fig:CDW} (Color online) Real-space charge and spin
    correlations relative to the central site for (a), (b) the Hubbard model
    ($\alpha=0$), and (c), (d) a long-range interaction ($\alpha=1.23$). Here, $\lambda = 0$, $U/t =
    3.5$, $L=15$.}
\end{figure}

To illustrate this competition, we show in Fig.~\ref{fig:CDW} the real-space
charge-charge correlation function
\begin{equation}
  N(\bm{r})   = \langle \hat{n}_{\bm{r}}  \hat{n}_{\bm{0}}\rangle -  \langle \hat{n}_{\bm{r}}  \rangle  
  \langle \hat{n}_{\bm{0}}\rangle \,,
\end{equation}  
and the spin-spin correlation function
\begin{equation}
  S(\bm{r})   = \langle \bm{\hat{S}}_{\bm{r}} \cdot
  \bm{\hat{S}}_{\bm{0}}\rangle \,.
\end{equation}  
The results are for $\lambda=0$ and $U<U_\text{c}$, corresponding to the
semimetallic phase. For $\alpha=0$ (Hubbard interaction),  $N(\bm{r})$ is
slightly suppressed around the origin with respect to the noninteracting
system.  This is typical of a liquid phase with contact interactions where
charges avoid each other at short distances. 

In the case of a long-range interaction, $\alpha =1.23$, we find enhanced
short-range charge correlations. At the same time, on going from $\alpha=0$
to $\alpha=1.23$, we observe a significant suppression of spin
correlations. These numerical data highlight the competition between charge
and spin order, and hence support the explanation of the shift of $\Uc$ in
terms of competing orders. Interestingly, even for $\alpha=1.23$---where
charge and spin correlations are nearly degenerate in the classical limit---we do
not find a stable charge-ordered phase, but a direct transition from the SM
to the AFMI phase.

\subsection{Magnetic phase transition at $\lambda/t=0.2$}\label{sec:TBI-Mott}

\begin{figure}[t]
  \includegraphics[width=0.45\textwidth]{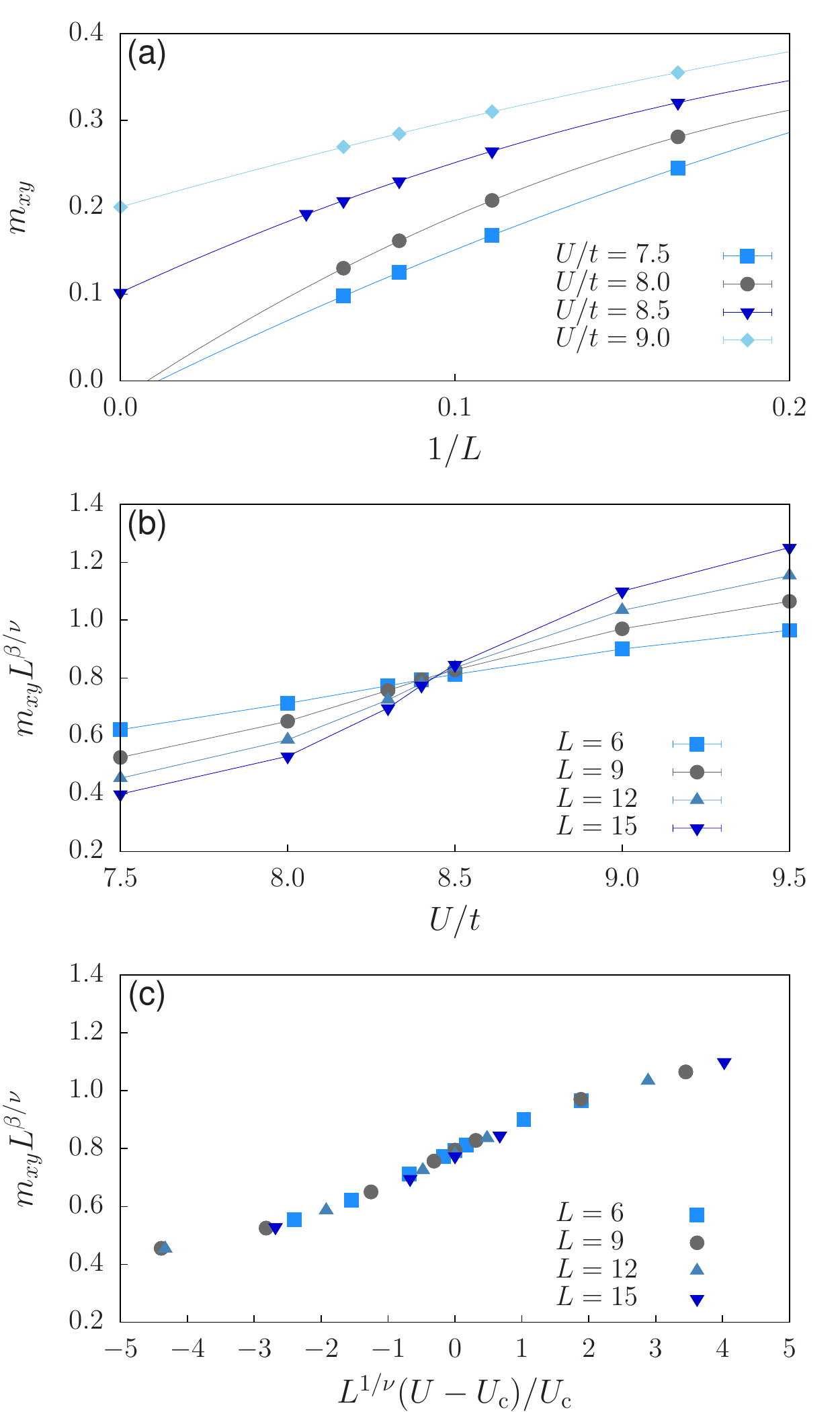}
  \caption{\label{fig:scaling_lambda0.2} (Color online) QSHI--AFMI quantum
    phase transition. (a) Finite-size
    scaling of the magnetization $m_{xy}$ using quadratic fits. (b) Scaling intersection using
    the critical exponents of the 3D XY model
    \cite{PhysRevB.74.144506,Ho.Me.La.We.Mu.As.12}. (c) Scaling collapse
    using  $U_\text{c}/t=8.4$. Here, $\lambda/t=0.2$,
    $\alpha=1$.}
\end{figure}

Similar to the KMH model \cite{RaHu10,Ho.Me.La.We.Mu.As.12}, the presence of
spin-orbit coupling is expected to allow  long-range magnetic order only in
the transverse spin direction. To determine the phase boundary, we therefore
measure the transverse spin correlation function~\footnote{For the $SU(2)$
symmetric case ($\lambda=0$), we have the relation
$S_{\alpha\beta}(\bm{i}-\bm{j})=({3}/{4}) S^{\pm}_{\alpha\beta}(\bm{i}-\bm{j})$.}
\begin{equation}\label{eq:S+S-}
  S^{\pm}_{\alpha\beta}(\bm{i}-\bm{j}) = \las S^+_{\bm{i}}  S^-_{\bm{j}} +
  S^-_{\bm{i}}  S^+_{\bm{j}}\ras 
\end{equation}
from which we obtain the structure factor
\begin{equation}\label{eq:SAF}
 \Sxy
 = \frac{1}{L^2}\sum_\alpha\sum_{\bm{i},\bm{j}} 
   S^{\pm}_{\alpha\alpha}(\bm{i}-\bm{j})
\end{equation}
and the transverse magnetization 
\begin{equation}\label{eq:mAF}
  m_{xy}=\sqrt{\Sxy /N}\,.
\end{equation}

Figure~\ref{fig:scaling_lambda0.2}(a) shows a finite-size scaling of $m_{xy}$ for
different values of $U/t$ for the KMC model with $\alpha=1$. The fits of the
data to second-order polynomials suggest that the critical point is located
in the range $U_\text{c}/t\in[8,8.5]$, compared to the value $U_\text{c}/t=5.70(3)$ found for
the KMH model at the same spin-orbit coupling $\lambda/t=0.2$.  The enhanced
critical value compared to Hubbard case can again be attributed to the
competition between charge and spin correlations, see
Sec.~\ref{sec:comp}. Similar to the KMH model \cite{Ho.Me.La.We.Mu.As.12}, we find no magnetic order
in the spin-$z$ direction over the whole range of interactions considered.

We can further test if the assumption of 3D XY universality, as previously demonstrated for
the analogous transition in the KMH model \cite{Ho.Me.La.We.Mu.As.12}, is consistent with our numerical data.
Figure~\ref{fig:scaling_lambda0.2}(b) shows the quantity
$L^{\beta/\nu} m_{xy}$ as a function of $U$ for different system sizes,
taking the critical exponents $z=1$, $\nu=0.6717(1)$ and $\beta=0.3486(1)$ of
the 3D XY model \cite{PhysRevB.74.144506}. We find an intersection of curves
for different system sizes at a value of $U_\text{c}/t=8.4(1)$, compatible
with Fig.~\ref{fig:scaling_lambda0.2}(a). In contrast to $\lambda=0$, we do
not observe logarithmic corrections to scaling as a result of the
long-range interaction. Nevertheless, the large critical value of the
transition renders simulations on large systems very demanding.

Taking $U_\text{c}/t=8.4$, we can produce a satisfactory scaling collapse in
Fig.~\ref{fig:scaling_lambda0.2}(c). The consistency between the onset of the
magnetization and the scaling intersection and collapse using the critical
exponents of the 3D XY model suggests
that the universality class of the transition is not changed by the long-range interaction.
In particular, the quality of the intersection and the data collapse in
Fig.~\ref{fig:scaling_lambda0.2} is very similar to that for the KMH model
\cite{Ho.Me.La.We.Mu.As.12}. We attribute the insensitivity to the nonlocal
part of the interaction to the fact that the magnetic excitons (corresponding
to particle-hole pairs) involved in the transition are charge neutral, and therefore
not affected by modifications of the potential.

\begin{figure}[t]
  \includegraphics[width=0.45\textwidth]{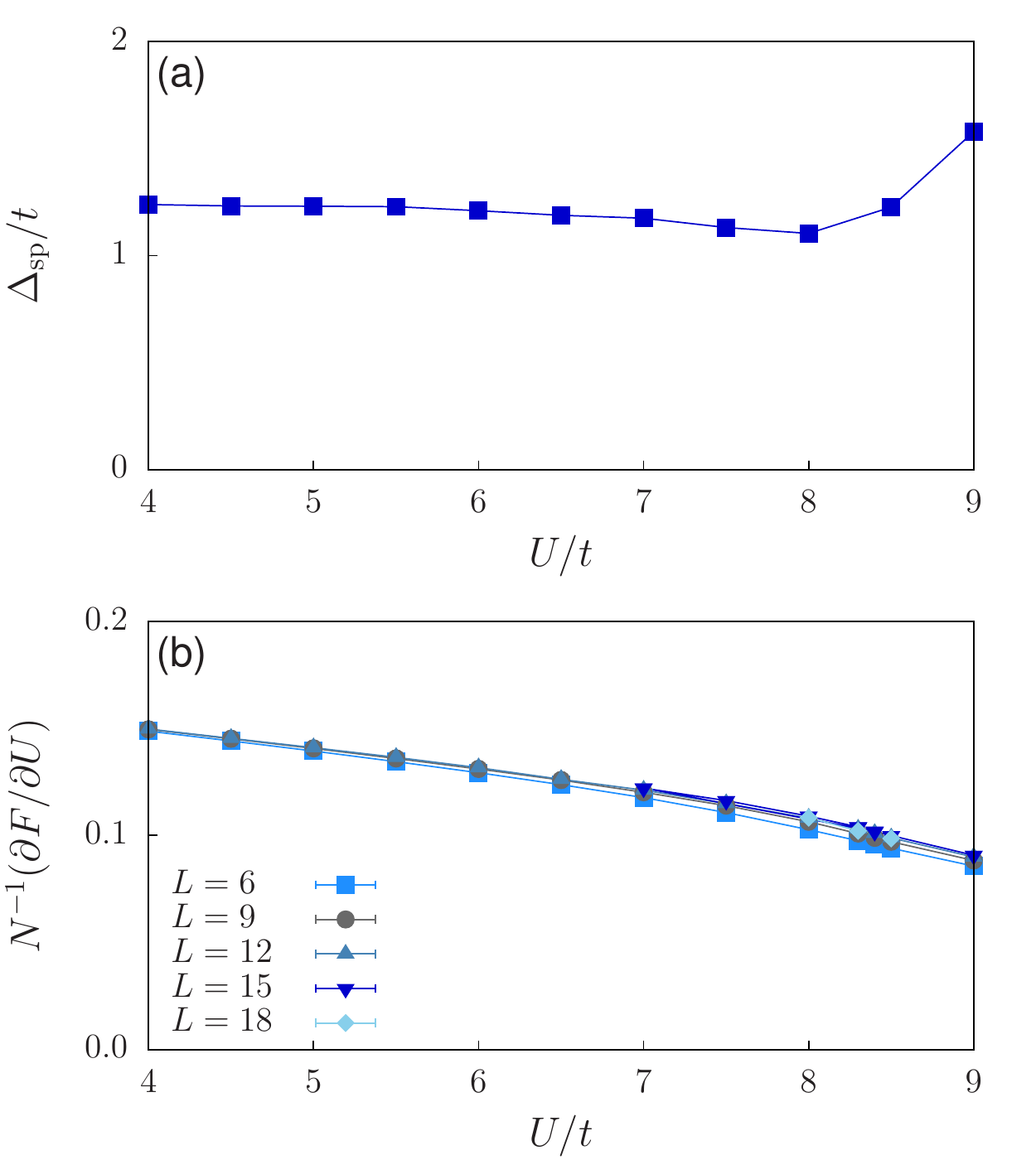}
  \caption{\label{fig:spgap_lambda0.2} (Color online) (a) Finite-size
    extrapolated single-particle gap. (b) Expectation value of the interaction
    term, corresponding to the derivative of the free energy with respect to
    $U$. Here, $\lambda/t=0.2$, $\alpha=1$.}
\end{figure}

\subsection{Absence of an intermediate phase}

The shift of the phase boundary for the magnetic phase transition in the KMC
model to significantly larger values of $U/t$ provides room for the QSH* phase
predicted to emerge from the interplay of strong spin-orbit coupling and
strong electron-electron interaction in \chem{Na_2IrO_3} \cite{PhysRevLett.108.046401}.
In particular, the QSH phase undergoes a transition to the QSH* phase 
at sufficiently large values of the spin-orbit coupling
upon increasing the Hubbard interaction  \cite{PhysRevLett.108.046401}.  While the model
for \chem{Na_2IrO_3} includes a Rashba spin-orbit term that does not conserve
spin \cite{Irridates-Nagaosa,PhysRevLett.108.046401}, mean-field calculations
suggest that such an interaction is not essential for the existence of the
QSH* phase, and that this phase could also exist in the KMH model \cite{Fieteprivate}. Its
absence in numerical results for the KMH model may therefore
be due to the onset of magnetic order already at intermediate values
of $U/t$. (Similar to the QSH phase, the QSH* phase relies on time-reversal
symmetry, and can therefore not coexist with magnetism.) Because the onset of
magnetic order is shifted to stronger interactions, the KMC model can in
principle provide a more favorable setting to observe this exotic phase.

Because the QSH* phase is not adiabatically connected to the QSH phase of the
KM model, we expect this phase to manifest itself in terms of an additional phase
transition. While a closing of the single-particle gap is not generally
necessary in correlated systems, we still expect such a transition to leave a
signature in the evolution of the gap with increasing $U$. However, the
results for the single-particle gap of the KMC model shown in
Fig.~\ref{fig:spgap_lambda0.2}(a) are qualitatively the same as for the KMH
model, and can be reproduced at the mean-field level \cite{Ho.Me.La.We.Mu.As.12}. The single-particle gap
remains nonzero throughout the QSH phase, and shows a single cusp at
the critical point of the magnetic transition. Similarly, and as in the case
of $\lambda=0$, the free-energy derivative with respect to the interaction 
shows a continuous evolution as a function of $U$, see Fig.~\ref{fig:spgap_lambda0.2}(b).
Finally, because the QSH* phase is expected
to be located between the QSH and the MI phase, it would change the
universality class of the magnetic transition, the latter being a QSH*-MI
transition instead of a QSH-MI transition. The scaling collapse obtained with
the 3D XY critical exponents in Fig.~\ref{fig:scaling_lambda0.2}(c) hence
contradicts the existence of an intermediate phase. Finally, the response to
$\pi$ fluxes could be used to measure the $Z_2$ topological invariant as a
function of $U$ \cite{As.Be.Ho.2012}.

\vspace*{1em}
\section{Conclusions}\label{sec:conclusions}

We have studied the Kane-Mele model with long-range Coulomb interaction using
an auxiliary-field quantum Monte Carlo method. The phase diagram shows
the same phases and phase transitions as for the Kane-Mele-Hubbard
model, namely a semimetal, a quantum spin Hall phase, and an
antiferromagnetic Mott insulator. Most notably, the magnetic transition is
shifted to significantly larger onsite interactions (in addition to the
nonlocal part) compared to a Hubbard interaction. This shift can be
understood as originating from the
competition between charge and spin order, with charge fluctuations being
enhanced by the nonlocal interactions.

The phase transitions between the semimetal and the antiferromagnetic
insulator in the absence of spin-orbit coupling, and between the quantum spin
Hall insulator and the antiferromagnetic insulator in the presence of
spin-orbit coupling, were analyzed with regard to the critical behavior. In
both cases, the critical exponents appear to be the same as for the
Hubbard interaction, namely those of the Gross-Neveu and the 3D XY
universality class, respectively. This observation agrees with analytical
findings regarding the marginal irrelevance of the long-range
interaction. Compared to the case of a Hubbard interaction, the problem with
long-range interactions is more challenging. Consequently, the finite-size
extrapolations and critical values are less accurate.

Finally, we did not find any evidence for additional phases. Our results
suggest that apart from quantitative differences, the Hubbard repulsion
captures the essential physics associated with strong
correlations. Unfortunately, because of a minus-sign problem, our method
cannot be applied to models with dominant nearest or next-nearest neighbor
interactions which may support additional symmetry-breaking phases \cite{RaQiHo08}.

{\begin{acknowledgments}%
We are grateful to the J\"ulich Supercomputing Centre for computer time, and
acknowledge financial support from the DFG Grant Nos. AS120/9-1 and
Ho 4489/2-1 (FOR 1807). We further thank A. R\"uegg, G. Fiete and L. Fritz
for valuable discussions.
\end{acknowledgments}}


\begin{thebibliography}{50}%
\makeatletter
\providecommand \@ifxundefined [1]{%
 \@ifx{#1\undefined}
}%
\providecommand \@ifnum [1]{%
 \ifnum #1\expandafter \@firstoftwo
 \else \expandafter \@secondoftwo
 \fi
}%
\providecommand \@ifx [1]{%
 \ifx #1\expandafter \@firstoftwo
 \else \expandafter \@secondoftwo
 \fi
}%
\providecommand \natexlab [1]{#1}%
\providecommand \enquote  [1]{``#1''}%
\providecommand \bibnamefont  [1]{#1}%
\providecommand \bibfnamefont [1]{#1}%
\providecommand \citenamefont [1]{#1}%
\providecommand \href@noop [0]{\@secondoftwo}%
\providecommand \href [0]{\begingroup \@sanitize@url \@href}%
\providecommand \@href[1]{\@@startlink{#1}\@@href}%
\providecommand \@@href[1]{\endgroup#1\@@endlink}%
\providecommand \@sanitize@url [0]{\catcode `\\12\catcode `\$12\catcode
  `\&12\catcode `\#12\catcode `\^12\catcode `\_12\catcode `\%12\relax}%
\providecommand \@@startlink[1]{}%
\providecommand \@@endlink[0]{}%
\providecommand \url  [0]{\begingroup\@sanitize@url \@url }%
\providecommand \@url [1]{\endgroup\@href {#1}{\urlprefix }}%
\providecommand \urlprefix  [0]{URL }%
\providecommand \Eprint [0]{\href }%
\providecommand \doibase [0]{http://dx.doi.org/}%
\providecommand \selectlanguage [0]{\@gobble}%
\providecommand \bibinfo  [0]{\@secondoftwo}%
\providecommand \bibfield  [0]{\@secondoftwo}%
\providecommand \translation [1]{[#1]}%
\providecommand \BibitemOpen [0]{}%
\providecommand \bibitemStop [0]{}%
\providecommand \bibitemNoStop [0]{.\EOS\space}%
\providecommand \EOS [0]{\spacefactor3000\relax}%
\providecommand \BibitemShut  [1]{\csname bibitem#1\endcsname}%
\let\auto@bib@innerbib\@empty
\bibitem [{\citenamefont {Novoselov}\ \emph {et~al.}(2005)\citenamefont
  {Novoselov}, \citenamefont {Geim}, \citenamefont {Morozov}, \citenamefont
  {Jiang}, \citenamefont {Katsnelson}, \citenamefont {Grigorieva},
  \citenamefont {Dubonos},\ and\ \citenamefont {Firsov}}]{Novoselov05}%
  \BibitemOpen
  \bibfield  {author} {\bibinfo {author} {\bibfnamefont {K.~S.}\ \bibnamefont
  {Novoselov}}, \bibinfo {author} {\bibfnamefont {A.~K.}\ \bibnamefont {Geim}},
  \bibinfo {author} {\bibfnamefont {S.~V.}\ \bibnamefont {Morozov}}, \bibinfo
  {author} {\bibfnamefont {D.}~\bibnamefont {Jiang}}, \bibinfo {author}
  {\bibfnamefont {M.~I.}\ \bibnamefont {Katsnelson}}, \bibinfo {author}
  {\bibfnamefont {I.~V.}\ \bibnamefont {Grigorieva}}, \bibinfo {author}
  {\bibfnamefont {S.~V.}\ \bibnamefont {Dubonos}}, \ and\ \bibinfo {author}
  {\bibfnamefont {A.~A.}\ \bibnamefont {Firsov}},\ }\href@noop {} {\bibfield
  {journal} {\bibinfo  {journal} {Nature}\ }\textbf {\bibinfo {volume} {438}},\
  \bibinfo {pages} {197} (\bibinfo {year} {2005})}\BibitemShut {NoStop}%
\bibitem [{\citenamefont {Kane}\ and\ \citenamefont
  {Mele}(2005{\natexlab{a}})}]{KaMe05b}%
  \BibitemOpen
  \bibfield  {author} {\bibinfo {author} {\bibfnamefont {C.~L.}\ \bibnamefont
  {Kane}}\ and\ \bibinfo {author} {\bibfnamefont {E.~J.}\ \bibnamefont
  {Mele}},\ }\href {\doibase 10.1103/PhysRevLett.95.226801} {\bibfield
  {journal} {\bibinfo  {journal} {Phys. Rev. Lett.}\ }\textbf {\bibinfo
  {volume} {95}},\ \bibinfo {pages} {226801} (\bibinfo {year}
  {2005}{\natexlab{a}})}\BibitemShut {NoStop}%
\bibitem [{\citenamefont {Raghu}\ \emph {et~al.}(2008)\citenamefont {Raghu},
  \citenamefont {Qi}, \citenamefont {Honerkamp},\ and\ \citenamefont
  {Zhang}}]{RaQiHo08}%
  \BibitemOpen
  \bibfield  {author} {\bibinfo {author} {\bibfnamefont {S.}~\bibnamefont
  {Raghu}}, \bibinfo {author} {\bibfnamefont {X.}~\bibnamefont {Qi}}, \bibinfo
  {author} {\bibfnamefont {C.}~\bibnamefont {Honerkamp}}, \ and\ \bibinfo
  {author} {\bibfnamefont {S.}~\bibnamefont {Zhang}},\ }\href {\doibase
  10.1103/PhysRevLett.100.156401} {\bibfield  {journal} {\bibinfo  {journal}
  {Phys. Rev. Lett.}\ }\textbf {\bibinfo {volume} {100}},\ \bibinfo {pages}
  {156401} (\bibinfo {year} {2008})}\BibitemShut {NoStop}%
\bibitem [{\citenamefont {Daghofer}\ and\ \citenamefont
  {Hohenadler}(2014)}]{PhysRevB.89.035103}%
  \BibitemOpen
  \bibfield  {author} {\bibinfo {author} {\bibfnamefont {M.}~\bibnamefont
  {Daghofer}}\ and\ \bibinfo {author} {\bibfnamefont {M.}~\bibnamefont
  {Hohenadler}},\ }\href {\doibase 10.1103/PhysRevB.89.035103} {\bibfield
  {journal} {\bibinfo  {journal} {Phys. Rev. B}\ }\textbf {\bibinfo {volume}
  {89}},\ \bibinfo {pages} {035103} (\bibinfo {year} {2014})}\BibitemShut
  {NoStop}%
\bibitem [{\citenamefont {Garc\'{i}a-Mart\'{i}nez}\ \emph
  {et~al.}(2013)\citenamefont {Garc\'{i}a-Mart\'{i}nez}, \citenamefont
  {Grushin}, \citenamefont {Neupert}, \citenamefont {Valenzuela},\ and\
  \citenamefont {Castro}}]{PhysRevB.88.245123}%
  \BibitemOpen
  \bibfield  {author} {\bibinfo {author} {\bibfnamefont {N.~A.}\ \bibnamefont
  {Garc\'{i}a-Mart\'{i}nez}}, \bibinfo {author} {\bibfnamefont {A.~G.}\
  \bibnamefont {Grushin}}, \bibinfo {author} {\bibfnamefont {T.}~\bibnamefont
  {Neupert}}, \bibinfo {author} {\bibfnamefont {B.}~\bibnamefont {Valenzuela}},
  \ and\ \bibinfo {author} {\bibfnamefont {E.~V.}\ \bibnamefont {Castro}},\
  }\href {\doibase 10.1103/PhysRevB.88.245123} {\bibfield  {journal} {\bibinfo
  {journal} {Phys. Rev. B}\ }\textbf {\bibinfo {volume} {88}},\ \bibinfo
  {pages} {245123} (\bibinfo {year} {2013})}\BibitemShut {NoStop}%
\bibitem [{\citenamefont {Duric}\ \emph {et~al.}(2014)\citenamefont {Duric},
  \citenamefont {Chancellor},\ and\ \citenamefont
  {Herbut}}]{PhysRevB.89.165123}%
  \BibitemOpen
  \bibfield  {author} {\bibinfo {author} {\bibfnamefont {T.}~\bibnamefont
  {Duric}}, \bibinfo {author} {\bibfnamefont {N.}~\bibnamefont {Chancellor}}, \
  and\ \bibinfo {author} {\bibfnamefont {I.~F.}\ \bibnamefont {Herbut}},\
  }\href {\doibase 10.1103/PhysRevB.89.165123} {\bibfield  {journal} {\bibinfo
  {journal} {Phys. Rev. B}\ }\textbf {\bibinfo {volume} {89}},\ \bibinfo
  {pages} {165123} (\bibinfo {year} {2014})}\BibitemShut {NoStop}%
\bibitem [{\citenamefont {Meng}\ \emph {et~al.}(2010)\citenamefont {Meng},
  \citenamefont {Lang}, \citenamefont {Wessel}, \citenamefont {Assaad},\ and\
  \citenamefont {Muramatsu}}]{Meng10}%
  \BibitemOpen
  \bibfield  {author} {\bibinfo {author} {\bibfnamefont {Z.~Y.}\ \bibnamefont
  {Meng}}, \bibinfo {author} {\bibfnamefont {T.~C.}\ \bibnamefont {Lang}},
  \bibinfo {author} {\bibfnamefont {S.}~\bibnamefont {Wessel}}, \bibinfo
  {author} {\bibfnamefont {F.~F.}\ \bibnamefont {Assaad}}, \ and\ \bibinfo
  {author} {\bibfnamefont {A.}~\bibnamefont {Muramatsu}},\ }\href@noop {}
  {\bibfield  {journal} {\bibinfo  {journal} {Nature}\ }\textbf {\bibinfo
  {volume} {464}},\ \bibinfo {pages} {847} (\bibinfo {year}
  {2010})}\BibitemShut {NoStop}%
\bibitem [{\citenamefont {Chen}\ \emph {et~al.}(2012)\citenamefont {Chen},
  \citenamefont {Gu}, \citenamefont {Liu},\ and\ \citenamefont
  {Wen}}]{Chen2012}%
  \BibitemOpen
  \bibfield  {author} {\bibinfo {author} {\bibfnamefont {X.}~\bibnamefont
  {Chen}}, \bibinfo {author} {\bibfnamefont {Z.-C.}\ \bibnamefont {Gu}},
  \bibinfo {author} {\bibfnamefont {Z.-X.}\ \bibnamefont {Liu}}, \ and\
  \bibinfo {author} {\bibfnamefont {X.-G.}\ \bibnamefont {Wen}},\ }\href@noop
  {} {\bibfield  {journal} {\bibinfo  {journal} {Science}\ }\textbf {\bibinfo
  {volume} {338}},\ \bibinfo {pages} {1604} (\bibinfo {year}
  {2012})}\BibitemShut {NoStop}%
\bibitem [{\citenamefont {Sorella}\ \emph {et~al.}(2012)\citenamefont
  {Sorella}, \citenamefont {Otsuka},\ and\ \citenamefont {Yunoki}}]{So.Ot.Yu.}%
  \BibitemOpen
  \bibfield  {author} {\bibinfo {author} {\bibfnamefont {S.}~\bibnamefont
  {Sorella}}, \bibinfo {author} {\bibfnamefont {Y.}~\bibnamefont {Otsuka}}, \
  and\ \bibinfo {author} {\bibfnamefont {S.}~\bibnamefont {Yunoki}},\
  }\href@noop {} {\bibfield  {journal} {\bibinfo  {journal} {Sci. Rep.}\
  }\textbf {\bibinfo {volume} {2}},\ \bibinfo {pages} {992} (\bibinfo {year}
  {2012})}\BibitemShut {NoStop}%
\bibitem [{\citenamefont {Assaad}\ and\ \citenamefont
  {Herbut}(2013)}]{Assaad13}%
  \BibitemOpen
  \bibfield  {author} {\bibinfo {author} {\bibfnamefont {F.~F.}\ \bibnamefont
  {Assaad}}\ and\ \bibinfo {author} {\bibfnamefont {I.~F.}\ \bibnamefont
  {Herbut}},\ }\href {\doibase 10.1103/PhysRevX.3.031010} {\bibfield  {journal}
  {\bibinfo  {journal} {Phys. Rev. X}\ }\textbf {\bibinfo {volume} {3}},\
  \bibinfo {pages} {031010} (\bibinfo {year} {2013})}\BibitemShut {NoStop}%
\bibitem [{\citenamefont {Clark}(2013)}]{Clark2013}%
  \BibitemOpen
  \bibfield  {author} {\bibinfo {author} {\bibfnamefont {B.~K.}\ \bibnamefont
  {Clark}},\ }\href@noop {} {\bibfield  {journal} {\bibinfo  {journal}
  {arXiv:1305.0278}\ } (\bibinfo {year} {2013})}\BibitemShut {NoStop}%
\bibitem [{\citenamefont {Shitade}\ \emph {et~al.}(2009)\citenamefont
  {Shitade}, \citenamefont {Katsura}, \citenamefont {Kunes}, \citenamefont
  {Qi}, \citenamefont {Zhang},\ and\ \citenamefont
  {Nagaosa}}]{Irridates-Nagaosa}%
  \BibitemOpen
  \bibfield  {author} {\bibinfo {author} {\bibfnamefont {A.}~\bibnamefont
  {Shitade}}, \bibinfo {author} {\bibfnamefont {H.}~\bibnamefont {Katsura}},
  \bibinfo {author} {\bibfnamefont {J.}~\bibnamefont {Kunes}}, \bibinfo
  {author} {\bibfnamefont {X.-L.}\ \bibnamefont {Qi}}, \bibinfo {author}
  {\bibfnamefont {S.-C.}\ \bibnamefont {Zhang}}, \ and\ \bibinfo {author}
  {\bibfnamefont {N.}~\bibnamefont {Nagaosa}},\ }\href@noop {} {\bibfield
  {journal} {\bibinfo  {journal} {Phys. Rev. Lett.}\ }\textbf {\bibinfo
  {volume} {102}},\ \bibinfo {pages} {256403} (\bibinfo {year}
  {2009})}\BibitemShut {NoStop}%
\bibitem [{\citenamefont {R\"uegg}\ and\ \citenamefont
  {Fiete}(2012)}]{PhysRevLett.108.046401}%
  \BibitemOpen
  \bibfield  {author} {\bibinfo {author} {\bibfnamefont {A.}~\bibnamefont
  {R\"uegg}}\ and\ \bibinfo {author} {\bibfnamefont {G.~A.}\ \bibnamefont
  {Fiete}},\ }\href {\doibase 10.1103/PhysRevLett.108.046401} {\bibfield
  {journal} {\bibinfo  {journal} {Phys. Rev. Lett.}\ }\textbf {\bibinfo
  {volume} {108}},\ \bibinfo {pages} {046401} (\bibinfo {year}
  {2012})}\BibitemShut {NoStop}%
\bibitem [{\citenamefont {Hubbard}(1963)}]{Hu63}%
  \BibitemOpen
  \bibfield  {author} {\bibinfo {author} {\bibfnamefont {J.}~\bibnamefont
  {Hubbard}},\ }\href@noop {} {\bibfield  {journal} {\bibinfo  {journal} {Proc.
  R. Soc. London}\ }\textbf {\bibinfo {volume} {276}},\ \bibinfo {pages} {238}
  (\bibinfo {year} {1963})}\BibitemShut {NoStop}%
\bibitem [{\citenamefont {Sorella}\ and\ \citenamefont
  {Tosatti}(1992)}]{Sorella92}%
  \BibitemOpen
  \bibfield  {author} {\bibinfo {author} {\bibfnamefont {S.}~\bibnamefont
  {Sorella}}\ and\ \bibinfo {author} {\bibfnamefont {E.}~\bibnamefont
  {Tosatti}},\ }\href@noop {} {\bibfield  {journal} {\bibinfo  {journal}
  {Europhys. Lett.}\ }\textbf {\bibinfo {volume} {19}},\ \bibinfo {pages} {699}
  (\bibinfo {year} {1992})}\BibitemShut {NoStop}%
\bibitem [{\citenamefont {Paiva}\ \emph {et~al.}(2005)\citenamefont {Paiva},
  \citenamefont {Scalettar}, \citenamefont {Zheng}, \citenamefont {Singh},\
  and\ \citenamefont {Oitmaa}}]{PhysRevB.72.085123}%
  \BibitemOpen
  \bibfield  {author} {\bibinfo {author} {\bibfnamefont {T.}~\bibnamefont
  {Paiva}}, \bibinfo {author} {\bibfnamefont {R.~T.}\ \bibnamefont
  {Scalettar}}, \bibinfo {author} {\bibfnamefont {W.}~\bibnamefont {Zheng}},
  \bibinfo {author} {\bibfnamefont {R.~R.~P.}\ \bibnamefont {Singh}}, \ and\
  \bibinfo {author} {\bibfnamefont {J.}~\bibnamefont {Oitmaa}},\ }\href
  {\doibase 10.1103/PhysRevB.72.085123} {\bibfield  {journal} {\bibinfo
  {journal} {Phys. Rev. B}\ }\textbf {\bibinfo {volume} {72}},\ \bibinfo
  {pages} {085123} (\bibinfo {year} {2005})}\BibitemShut {NoStop}%
\bibitem [{\citenamefont {Hohenadler}\ \emph {et~al.}(2011)\citenamefont
  {Hohenadler}, \citenamefont {Lang},\ and\ \citenamefont
  {Assaad}}]{Hohenadler10}%
  \BibitemOpen
  \bibfield  {author} {\bibinfo {author} {\bibfnamefont {M.}~\bibnamefont
  {Hohenadler}}, \bibinfo {author} {\bibfnamefont {T.~C.}\ \bibnamefont
  {Lang}}, \ and\ \bibinfo {author} {\bibfnamefont {F.~F.}\ \bibnamefont
  {Assaad}},\ }\href@noop {} {\bibfield  {journal} {\bibinfo  {journal} {Phys.
  Rev. Lett.}\ }\textbf {\bibinfo {volume} {106}},\ \bibinfo {pages} {100403}
  (\bibinfo {year} {2011})}\BibitemShut {NoStop}%
\bibitem [{\citenamefont {Zheng}\ \emph {et~al.}(2011)\citenamefont {Zheng},
  \citenamefont {Zhang},\ and\ \citenamefont {Wu}}]{Zh.Wu.Zh.11}%
  \BibitemOpen
  \bibfield  {author} {\bibinfo {author} {\bibfnamefont {D.}~\bibnamefont
  {Zheng}}, \bibinfo {author} {\bibfnamefont {G.-M.}\ \bibnamefont {Zhang}}, \
  and\ \bibinfo {author} {\bibfnamefont {C.}~\bibnamefont {Wu}},\ }\href
  {\doibase 10.1103/PhysRevB.84.205121} {\bibfield  {journal} {\bibinfo
  {journal} {Phys. Rev. B}\ }\textbf {\bibinfo {volume} {84}},\ \bibinfo
  {pages} {205121} (\bibinfo {year} {2011})}\BibitemShut {NoStop}%
\bibitem [{\citenamefont {Rachel}\ and\ \citenamefont {{Le
  Hur}}(2010)}]{RaHu10}%
  \BibitemOpen
  \bibfield  {author} {\bibinfo {author} {\bibfnamefont {S.}~\bibnamefont
  {Rachel}}\ and\ \bibinfo {author} {\bibfnamefont {K.}~\bibnamefont {{Le
  Hur}}},\ }\href {\doibase 10.1103/PhysRevB.82.075106} {\bibfield  {journal}
  {\bibinfo  {journal} {Phys. Rev. B}\ }\textbf {\bibinfo {volume} {82}},\
  \bibinfo {pages} {075106} (\bibinfo {year} {2010})}\BibitemShut {NoStop}%
\bibitem [{\citenamefont {Hohenadler}\ and\ \citenamefont
  {Assaad}(2013)}]{HoAsreview2013}%
  \BibitemOpen
  \bibfield  {author} {\bibinfo {author} {\bibfnamefont {M.}~\bibnamefont
  {Hohenadler}}\ and\ \bibinfo {author} {\bibfnamefont {F.~F.}\ \bibnamefont
  {Assaad}},\ }\href {\doibase 10.1088/0953-8984/25/14/143201} {\bibfield
  {journal} {\bibinfo  {journal} {J. Phys.: Condens. Matter}\ }\textbf
  {\bibinfo {volume} {25}},\ \bibinfo {pages} {143201} (\bibinfo {year}
  {2013})}\BibitemShut {NoStop}%
\bibitem [{\citenamefont {{Castro Neto}}\ \emph {et~al.}(2009)\citenamefont
  {{Castro Neto}}, \citenamefont {Guinea}, \citenamefont {Peres}, \citenamefont
  {Novoselov},\ and\ \citenamefont {Geim}}]{Neto_rev}%
  \BibitemOpen
  \bibfield  {author} {\bibinfo {author} {\bibfnamefont {A.~H.}\ \bibnamefont
  {{Castro Neto}}}, \bibinfo {author} {\bibfnamefont {F.}~\bibnamefont
  {Guinea}}, \bibinfo {author} {\bibfnamefont {N.~M.~R.}\ \bibnamefont
  {Peres}}, \bibinfo {author} {\bibfnamefont {K.~S.}\ \bibnamefont
  {Novoselov}}, \ and\ \bibinfo {author} {\bibfnamefont {A.~K.}\ \bibnamefont
  {Geim}},\ }\href {\doibase 10.1103/RevModPhys.81.109} {\bibfield  {journal}
  {\bibinfo  {journal} {Rev. Mod. Phys.}\ }\textbf {\bibinfo {volume} {81}},\
  \bibinfo {eid} {109} (\bibinfo {year} {2009})}\BibitemShut {NoStop}%
\bibitem [{\citenamefont {Wallace}(1947)}]{Wallace47}%
  \BibitemOpen
  \bibfield  {author} {\bibinfo {author} {\bibfnamefont {P.~R.}\ \bibnamefont
  {Wallace}},\ }\href@noop {} {\bibfield  {journal} {\bibinfo  {journal} {Phys.
  Rev.}\ }\textbf {\bibinfo {volume} {71}},\ \bibinfo {pages} {622} (\bibinfo
  {year} {1947})}\BibitemShut {NoStop}%
\bibitem [{\citenamefont {Herbut}(2006)}]{PhysRevLett.97.146401}%
  \BibitemOpen
  \bibfield  {author} {\bibinfo {author} {\bibfnamefont {I.~F.}\ \bibnamefont
  {Herbut}},\ }\href {\doibase 10.1103/PhysRevLett.97.146401} {\bibfield
  {journal} {\bibinfo  {journal} {Phys. Rev. Lett.}\ }\textbf {\bibinfo
  {volume} {97}},\ \bibinfo {pages} {146401} (\bibinfo {year}
  {2006})}\BibitemShut {NoStop}%
\bibitem [{\citenamefont {Herbut}\ \emph
  {et~al.}(2009{\natexlab{a}})\citenamefont {Herbut}, \citenamefont
  {Juri\ifmmode \check{c}\else \v{c}\fi{}i\ifmmode~\acute{c}\else \'{c}\fi{}},\
  and\ \citenamefont {Vafek}}]{Herbut09a}%
  \BibitemOpen
  \bibfield  {author} {\bibinfo {author} {\bibfnamefont {I.~F.}\ \bibnamefont
  {Herbut}}, \bibinfo {author} {\bibfnamefont {V.}~\bibnamefont {Juri\ifmmode
  \check{c}\else \v{c}\fi{}i\ifmmode~\acute{c}\else \'{c}\fi{}}}, \ and\
  \bibinfo {author} {\bibfnamefont {O.}~\bibnamefont {Vafek}},\ }\href
  {\doibase 10.1103/PhysRevB.80.075432} {\bibfield  {journal} {\bibinfo
  {journal} {Phys. Rev. B}\ }\textbf {\bibinfo {volume} {80}},\ \bibinfo
  {pages} {075432} (\bibinfo {year} {2009}{\natexlab{a}})}\BibitemShut
  {NoStop}%
\bibitem [{\citenamefont {Janssen}\ and\ \citenamefont
  {Herbut}(2014)}]{PhysRevB.89.205403}%
  \BibitemOpen
  \bibfield  {author} {\bibinfo {author} {\bibfnamefont {L.}~\bibnamefont
  {Janssen}}\ and\ \bibinfo {author} {\bibfnamefont {I.~F.}\ \bibnamefont
  {Herbut}},\ }\href {\doibase 10.1103/PhysRevB.89.205403} {\bibfield
  {journal} {\bibinfo  {journal} {Phys. Rev. B}\ }\textbf {\bibinfo {volume}
  {89}},\ \bibinfo {pages} {205403} (\bibinfo {year} {2014})}\BibitemShut
  {NoStop}%
\bibitem [{\citenamefont {Gonzalez}\ \emph {et~al.}(1994)\citenamefont
  {Gonzalez}, \citenamefont {Guinea},\ and\ \citenamefont
  {Vozmediano}}]{Gonzalez1994595}%
  \BibitemOpen
  \bibfield  {author} {\bibinfo {author} {\bibfnamefont {J.}~\bibnamefont
  {Gonzalez}}, \bibinfo {author} {\bibfnamefont {F.}~\bibnamefont {Guinea}}, \
  and\ \bibinfo {author} {\bibfnamefont {M.~A.~H.}\ \bibnamefont
  {Vozmediano}},\ }\href@noop {} {\bibfield  {journal} {\bibinfo  {journal}
  {Nuclear Physics B}\ }\textbf {\bibinfo {volume} {424}},\ \bibinfo {pages}
  {595 } (\bibinfo {year} {1994})}\BibitemShut {NoStop}%
\bibitem [{\citenamefont {Gonz\'alez}\ \emph {et~al.}(1999)\citenamefont
  {Gonz\'alez}, \citenamefont {Guinea},\ and\ \citenamefont
  {Vozmediano}}]{PhysRevB.59.R2474}%
  \BibitemOpen
  \bibfield  {author} {\bibinfo {author} {\bibfnamefont {J.}~\bibnamefont
  {Gonz\'alez}}, \bibinfo {author} {\bibfnamefont {F.}~\bibnamefont {Guinea}},
  \ and\ \bibinfo {author} {\bibfnamefont {M.~A.~H.}\ \bibnamefont
  {Vozmediano}},\ }\href {\doibase 10.1103/PhysRevB.59.R2474} {\bibfield
  {journal} {\bibinfo  {journal} {Phys. Rev. B}\ }\textbf {\bibinfo {volume}
  {59}},\ \bibinfo {pages} {R2474} (\bibinfo {year} {1999})}\BibitemShut
  {NoStop}%
\bibitem [{\citenamefont {Elias}\ \emph {et~al.}(2011)\citenamefont {Elias},
  \citenamefont {Gorbachev}, \citenamefont {Mayorov}, \citenamefont {Morozov},
  \citenamefont {Zhukov}, \citenamefont {Blake}, \citenamefont {Ponomarenko},
  \citenamefont {Grigorieva}, \citenamefont {Novoselov}, \citenamefont
  {Guinea},\ and\ \citenamefont {Geim}}]{Elias11}%
  \BibitemOpen
  \bibfield  {author} {\bibinfo {author} {\bibfnamefont {D.~C.}\ \bibnamefont
  {Elias}}, \bibinfo {author} {\bibfnamefont {R.~V.}\ \bibnamefont
  {Gorbachev}}, \bibinfo {author} {\bibfnamefont {A.~S.}\ \bibnamefont
  {Mayorov}}, \bibinfo {author} {\bibfnamefont {S.~V.}\ \bibnamefont
  {Morozov}}, \bibinfo {author} {\bibfnamefont {A.~A.}\ \bibnamefont {Zhukov}},
  \bibinfo {author} {\bibfnamefont {P.}~\bibnamefont {Blake}}, \bibinfo
  {author} {\bibfnamefont {L.~A.}\ \bibnamefont {Ponomarenko}}, \bibinfo
  {author} {\bibfnamefont {I.~V.}\ \bibnamefont {Grigorieva}}, \bibinfo
  {author} {\bibfnamefont {K.~S.}\ \bibnamefont {Novoselov}}, \bibinfo {author}
  {\bibfnamefont {F.}~\bibnamefont {Guinea}}, \ and\ \bibinfo {author}
  {\bibfnamefont {A.~K.}\ \bibnamefont {Geim}},\ }\href@noop {} {\bibfield
  {journal} {\bibinfo  {journal} {Nat. Phys.}\ }\textbf {\bibinfo {volume}
  {7}},\ \bibinfo {pages} {701} (\bibinfo {year} {2011})}\BibitemShut {NoStop}%
\bibitem [{\citenamefont {Juri\ifmmode \check{c}\else
  \v{c}\fi{}i\ifmmode~\acute{c}\else \'{c}\fi{}}\ \emph
  {et~al.}(2009)\citenamefont {Juri\ifmmode \check{c}\else
  \v{c}\fi{}i\ifmmode~\acute{c}\else \'{c}\fi{}}, \citenamefont {Herbut},\ and\
  \citenamefont {Semenoff}}]{PhysRevB.80.081405}%
  \BibitemOpen
  \bibfield  {author} {\bibinfo {author} {\bibfnamefont {V.}~\bibnamefont
  {Juri\ifmmode \check{c}\else \v{c}\fi{}i\ifmmode~\acute{c}\else \'{c}\fi{}}},
  \bibinfo {author} {\bibfnamefont {I.~F.}\ \bibnamefont {Herbut}}, \ and\
  \bibinfo {author} {\bibfnamefont {G.~W.}\ \bibnamefont {Semenoff}},\ }\href
  {\doibase 10.1103/PhysRevB.80.081405} {\bibfield  {journal} {\bibinfo
  {journal} {Phys. Rev. B}\ }\textbf {\bibinfo {volume} {80}},\ \bibinfo
  {pages} {081405} (\bibinfo {year} {2009})}\BibitemShut {NoStop}%
\bibitem [{\citenamefont {Hohenadler}\ \emph {et~al.}(2012)\citenamefont
  {Hohenadler}, \citenamefont {Meng}, \citenamefont {Lang}, \citenamefont
  {Wessel}, \citenamefont {Muramatsu},\ and\ \citenamefont
  {Assaad}}]{Ho.Me.La.We.Mu.As.12}%
  \BibitemOpen
  \bibfield  {author} {\bibinfo {author} {\bibfnamefont {M.}~\bibnamefont
  {Hohenadler}}, \bibinfo {author} {\bibfnamefont {Z.~Y.}\ \bibnamefont
  {Meng}}, \bibinfo {author} {\bibfnamefont {T.~C.}\ \bibnamefont {Lang}},
  \bibinfo {author} {\bibfnamefont {S.}~\bibnamefont {Wessel}}, \bibinfo
  {author} {\bibfnamefont {A.}~\bibnamefont {Muramatsu}}, \ and\ \bibinfo
  {author} {\bibfnamefont {F.~F.}\ \bibnamefont {Assaad}},\ }\href@noop {}
  {\bibfield  {journal} {\bibinfo  {journal} {Phys. Rev. B}\ }\textbf {\bibinfo
  {volume} {85}},\ \bibinfo {pages} {115132} (\bibinfo {year}
  {2012})}\BibitemShut {NoStop}%
\bibitem [{\citenamefont {Drut}\ and\ \citenamefont
  {L\"ahde}(2009)}]{PhysRevLett.102.026802}%
  \BibitemOpen
  \bibfield  {author} {\bibinfo {author} {\bibfnamefont {J.~E.}\ \bibnamefont
  {Drut}}\ and\ \bibinfo {author} {\bibfnamefont {T.~A.}\ \bibnamefont
  {L\"ahde}},\ }\href {\doibase 10.1103/PhysRevLett.102.026802} {\bibfield
  {journal} {\bibinfo  {journal} {Phys. Rev. Lett.}\ }\textbf {\bibinfo
  {volume} {102}},\ \bibinfo {pages} {026802} (\bibinfo {year}
  {2009})}\BibitemShut {NoStop}%
\bibitem [{\citenamefont {Sekine}\ \emph {et~al.}(2013)\citenamefont {Sekine},
  \citenamefont {Nakano}, \citenamefont {Araki},\ and\ \citenamefont
  {Nomura}}]{PhysRevB.87.165142}%
  \BibitemOpen
  \bibfield  {author} {\bibinfo {author} {\bibfnamefont {A.}~\bibnamefont
  {Sekine}}, \bibinfo {author} {\bibfnamefont {T.~Z.}\ \bibnamefont {Nakano}},
  \bibinfo {author} {\bibfnamefont {Y.}~\bibnamefont {Araki}}, \ and\ \bibinfo
  {author} {\bibfnamefont {K.}~\bibnamefont {Nomura}},\ }\href {\doibase
  10.1103/PhysRevB.87.165142} {\bibfield  {journal} {\bibinfo  {journal} {Phys.
  Rev. B}\ }\textbf {\bibinfo {volume} {87}},\ \bibinfo {pages} {165142}
  (\bibinfo {year} {2013})}\BibitemShut {NoStop}%
\bibitem [{\citenamefont {Araki}\ and\ \citenamefont
  {Kimura}(2013)}]{PhysRevB.87.205440}%
  \BibitemOpen
  \bibfield  {author} {\bibinfo {author} {\bibfnamefont {Y.}~\bibnamefont
  {Araki}}\ and\ \bibinfo {author} {\bibfnamefont {T.}~\bibnamefont {Kimura}},\
  }\href {\doibase 10.1103/PhysRevB.87.205440} {\bibfield  {journal} {\bibinfo
  {journal} {Phys. Rev. B}\ }\textbf {\bibinfo {volume} {87}},\ \bibinfo
  {pages} {205440} (\bibinfo {year} {2013})}\BibitemShut {NoStop}%
\bibitem [{\citenamefont {Sekine}\ and\ \citenamefont
  {Nomura}(2013)}]{Sekine13}%
  \BibitemOpen
  \bibfield  {author} {\bibinfo {author} {\bibfnamefont {A.}~\bibnamefont
  {Sekine}}\ and\ \bibinfo {author} {\bibfnamefont {K.}~\bibnamefont
  {Nomura}},\ }\href@noop {} {\bibfield  {journal} {\bibinfo  {journal}
  {arXiv:1309.1079}\ } (\bibinfo {year} {2013})}\BibitemShut {NoStop}%
\bibitem [{\citenamefont {Sekine}\ and\ \citenamefont
  {Nomura}(2014)}]{Sekine14}%
  \BibitemOpen
  \bibfield  {author} {\bibinfo {author} {\bibfnamefont {A.}~\bibnamefont
  {Sekine}}\ and\ \bibinfo {author} {\bibfnamefont {K.}~\bibnamefont
  {Nomura}},\ }\href@noop {} {\bibfield  {journal} {\bibinfo  {journal}
  {arXiv:1405.6932}\ } (\bibinfo {year} {2014})}\BibitemShut {NoStop}%
\bibitem [{\citenamefont {Kane}\ and\ \citenamefont
  {Mele}(2005{\natexlab{b}})}]{KaMe05a}%
  \BibitemOpen
  \bibfield  {author} {\bibinfo {author} {\bibfnamefont {C.~L.}\ \bibnamefont
  {Kane}}\ and\ \bibinfo {author} {\bibfnamefont {E.~J.}\ \bibnamefont
  {Mele}},\ }\href {\doibase 10.1103/PhysRevLett.95.146802} {\bibfield
  {journal} {\bibinfo  {journal} {Phys. Rev. Lett.}\ }\textbf {\bibinfo
  {volume} {95}},\ \bibinfo {pages} {146802} (\bibinfo {year}
  {2005}{\natexlab{b}})}\BibitemShut {NoStop}%
\bibitem [{\citenamefont {Ulybyshev}\ \emph {et~al.}(2013)\citenamefont
  {Ulybyshev}, \citenamefont {Buividovich}, \citenamefont {Katsnelson},\ and\
  \citenamefont {Polikarpov}}]{Ulybyshev2013}%
  \BibitemOpen
  \bibfield  {author} {\bibinfo {author} {\bibfnamefont {M.~V.}\ \bibnamefont
  {Ulybyshev}}, \bibinfo {author} {\bibfnamefont {P.~V.}\ \bibnamefont
  {Buividovich}}, \bibinfo {author} {\bibfnamefont {M.~I.}\ \bibnamefont
  {Katsnelson}}, \ and\ \bibinfo {author} {\bibfnamefont {M.~I.}\ \bibnamefont
  {Polikarpov}},\ }\href {\doibase 10.1103/PhysRevLett.111.056801} {\bibfield
  {journal} {\bibinfo  {journal} {Phys. Rev. Lett.}\ }\textbf {\bibinfo
  {volume} {111}},\ \bibinfo {pages} {056801} (\bibinfo {year}
  {2013})}\BibitemShut {NoStop}%
\bibitem [{\citenamefont {Scalettar}\ \emph {et~al.}(1987)\citenamefont
  {Scalettar}, \citenamefont {Scalapino}, \citenamefont {Sugar},\ and\
  \citenamefont {Toussaint}}]{Scalettar87}%
  \BibitemOpen
  \bibfield  {author} {\bibinfo {author} {\bibfnamefont {R.~T.}\ \bibnamefont
  {Scalettar}}, \bibinfo {author} {\bibfnamefont {D.~J.}\ \bibnamefont
  {Scalapino}}, \bibinfo {author} {\bibfnamefont {R.~L.}\ \bibnamefont
  {Sugar}}, \ and\ \bibinfo {author} {\bibfnamefont {D.}~\bibnamefont
  {Toussaint}},\ }\href {\doibase 10.1103/PhysRevB.36.8632} {\bibfield
  {journal} {\bibinfo  {journal} {Phys. Rev. B}\ }\textbf {\bibinfo {volume}
  {36}},\ \bibinfo {pages} {8632} (\bibinfo {year} {1987})}\BibitemShut
  {NoStop}%
\bibitem [{\citenamefont {{Assaad}}\ and\ \citenamefont
  {{Evertz}}(2008)}]{Assaad08_rev}%
  \BibitemOpen
  \bibfield  {author} {\bibinfo {author} {\bibfnamefont {F.~F.}\ \bibnamefont
  {{Assaad}}}\ and\ \bibinfo {author} {\bibfnamefont {H.~G.}\ \bibnamefont
  {{Evertz}}},\ }in\ \href@noop {} {\emph {\bibinfo {booktitle} {Computational
  Many Particle Physics}}},\ \bibinfo {series} {Lecture Notes in Physics},
  Vol.\ \bibinfo {volume} {739},\ \bibinfo {editor} {edited by\ \bibinfo
  {editor} {\bibfnamefont {H.}~\bibnamefont {{Fehske}}}, \bibinfo {editor}
  {\bibfnamefont {R.}~\bibnamefont {{Schneider}}}, \ and\ \bibinfo {editor}
  {\bibfnamefont {A.}~\bibnamefont {{Wei{\ss}e}}}}\ (\bibinfo  {publisher}
  {Springer Verlag},\ \bibinfo {address} {Berlin},\ \bibinfo {year} {2008})\
  p.\ \bibinfo {pages} {277}\BibitemShut {NoStop}%
\bibitem [{\citenamefont {Feldbacher}\ and\ \citenamefont
  {Assaad}(2001)}]{Feldbach00}%
  \BibitemOpen
  \bibfield  {author} {\bibinfo {author} {\bibfnamefont {M.}~\bibnamefont
  {Feldbacher}}\ and\ \bibinfo {author} {\bibfnamefont {F.~F.}\ \bibnamefont
  {Assaad}},\ }\href@noop {} {\bibfield  {journal} {\bibinfo  {journal} {Phys.
  Rev. B}\ }\textbf {\bibinfo {volume} {63}},\ \bibinfo {pages} {073105}
  (\bibinfo {year} {2001})}\BibitemShut {NoStop}%
\bibitem [{\citenamefont {Assaad}\ \emph {et~al.}(2013)\citenamefont {Assaad},
  \citenamefont {Bercx},\ and\ \citenamefont {Hohenadler}}]{As.Be.Ho.2012}%
  \BibitemOpen
  \bibfield  {author} {\bibinfo {author} {\bibfnamefont {F.~F.}\ \bibnamefont
  {Assaad}}, \bibinfo {author} {\bibfnamefont {M.}~\bibnamefont {Bercx}}, \
  and\ \bibinfo {author} {\bibfnamefont {M.}~\bibnamefont {Hohenadler}},\
  }\href@noop {} {\bibfield  {journal} {\bibinfo  {journal} {Phys. Rev. X}\
  }\textbf {\bibinfo {volume} {3}},\ \bibinfo {pages} {011015} (\bibinfo {year}
  {2013})}\BibitemShut {NoStop}%
\bibitem [{\citenamefont {Assaad}\ and\ \citenamefont
  {Hohenadler}(2013)}]{inside2013}%
  \BibitemOpen
  \bibfield  {author} {\bibinfo {author} {\bibfnamefont {F.~F.}\ \bibnamefont
  {Assaad}}\ and\ \bibinfo {author} {\bibfnamefont {M.}~\bibnamefont
  {Hohenadler}},\ }\href@noop {} {\bibfield  {journal} {\bibinfo  {journal}
  {inSIDE}\ }\textbf {\bibinfo {volume} {11}},\ \bibinfo {pages} {22} (\bibinfo
  {year} {2013})}\BibitemShut {NoStop}%
\bibitem [{\citenamefont {Herbut}\ \emph
  {et~al.}(2009{\natexlab{b}})\citenamefont {Herbut}, \citenamefont
  {Juri\ifmmode \check{c}\else \v{c}\fi{}i\ifmmode~\acute{c}\else \'{c}\fi{}},\
  and\ \citenamefont {Roy}}]{Herbut09}%
  \BibitemOpen
  \bibfield  {author} {\bibinfo {author} {\bibfnamefont {I.~F.}\ \bibnamefont
  {Herbut}}, \bibinfo {author} {\bibfnamefont {V.}~\bibnamefont {Juri\ifmmode
  \check{c}\else \v{c}\fi{}i\ifmmode~\acute{c}\else \'{c}\fi{}}}, \ and\
  \bibinfo {author} {\bibfnamefont {B.}~\bibnamefont {Roy}},\ }\href {\doibase
  10.1103/PhysRevB.79.085116} {\bibfield  {journal} {\bibinfo  {journal} {Phys.
  Rev. B}\ }\textbf {\bibinfo {volume} {79}},\ \bibinfo {pages} {085116}
  (\bibinfo {year} {2009}{\natexlab{b}})}\BibitemShut {NoStop}%
\bibitem [{\citenamefont {Chandrasekharan}\ and\ \citenamefont
  {Li}(2013)}]{PhysRevD.88.021701}%
  \BibitemOpen
  \bibfield  {author} {\bibinfo {author} {\bibfnamefont {S.}~\bibnamefont
  {Chandrasekharan}}\ and\ \bibinfo {author} {\bibfnamefont {A.}~\bibnamefont
  {Li}},\ }\href {\doibase 10.1103/PhysRevD.88.021701} {\bibfield  {journal}
  {\bibinfo  {journal} {Phys. Rev. D}\ }\textbf {\bibinfo {volume} {88}},\
  \bibinfo {pages} {021701} (\bibinfo {year} {2013})}\BibitemShut {NoStop}%
\bibitem [{\citenamefont {Roy}\ \emph {et~al.}(2013)\citenamefont {Roy},
  \citenamefont {Juri\ifmmode \check{c}\else \v{c}\fi{}i\ifmmode~\acute{c}\else
  \'{c}\fi{}},\ and\ \citenamefont {Herbut}}]{PhysRevB.87.041401}%
  \BibitemOpen
  \bibfield  {author} {\bibinfo {author} {\bibfnamefont {B.}~\bibnamefont
  {Roy}}, \bibinfo {author} {\bibfnamefont {V.}~\bibnamefont {Juri\ifmmode
  \check{c}\else \v{c}\fi{}i\ifmmode~\acute{c}\else \'{c}\fi{}}}, \ and\
  \bibinfo {author} {\bibfnamefont {I.~F.}\ \bibnamefont {Herbut}},\ }\href
  {\doibase 10.1103/PhysRevB.87.041401} {\bibfield  {journal} {\bibinfo
  {journal} {Phys. Rev. B}\ }\textbf {\bibinfo {volume} {87}},\ \bibinfo
  {pages} {041401} (\bibinfo {year} {2013})}\BibitemShut {NoStop}%
\bibitem [{\citenamefont {Sch\"uler}\ \emph {et~al.}(2013)\citenamefont
  {Sch\"uler}, \citenamefont {R\"osner}, \citenamefont {Wehling}, \citenamefont
  {Lichtenstein},\ and\ \citenamefont {Katsnelson}}]{Schuler13}%
  \BibitemOpen
  \bibfield  {author} {\bibinfo {author} {\bibfnamefont {M.}~\bibnamefont
  {Sch\"uler}}, \bibinfo {author} {\bibfnamefont {M.}~\bibnamefont {R\"osner}},
  \bibinfo {author} {\bibfnamefont {T.~O.}\ \bibnamefont {Wehling}}, \bibinfo
  {author} {\bibfnamefont {A.~I.}\ \bibnamefont {Lichtenstein}}, \ and\
  \bibinfo {author} {\bibfnamefont {M.~I.}\ \bibnamefont {Katsnelson}},\ }\href
  {\doibase 10.1103/PhysRevLett.111.036601} {\bibfield  {journal} {\bibinfo
  {journal} {Phys. Rev. Lett.}\ }\textbf {\bibinfo {volume} {111}},\ \bibinfo
  {pages} {036601} (\bibinfo {year} {2013})}\BibitemShut {NoStop}%
\bibitem [{\citenamefont {Campostrini}\ \emph {et~al.}(2006)\citenamefont
  {Campostrini}, \citenamefont {Hasenbusch}, \citenamefont {Pelissetto},\ and\
  \citenamefont {Vicari}}]{PhysRevB.74.144506}%
  \BibitemOpen
  \bibfield  {author} {\bibinfo {author} {\bibfnamefont {M.}~\bibnamefont
  {Campostrini}}, \bibinfo {author} {\bibfnamefont {M.}~\bibnamefont
  {Hasenbusch}}, \bibinfo {author} {\bibfnamefont {A.}~\bibnamefont
  {Pelissetto}}, \ and\ \bibinfo {author} {\bibfnamefont {E.}~\bibnamefont
  {Vicari}},\ }\href {\doibase 10.1103/PhysRevB.74.144506} {\bibfield
  {journal} {\bibinfo  {journal} {Phys. Rev. B}\ }\textbf {\bibinfo {volume}
  {74}},\ \bibinfo {pages} {144506} (\bibinfo {year} {2006})}\BibitemShut
  {NoStop}%
\bibitem [{Note1()}]{Note1}%
  \BibitemOpen
  \bibinfo {note} {For the $SU(2)$ symmetric case ($\lambda =0$), we have the
  relation $S_{\alpha \beta }(\protect \bm {i}-\protect \bm {j})=({3}/{4})
  S^{\pm }_{\alpha \beta }(\protect \bm {i}-\protect \bm {j})$.}\BibitemShut
  {Stop}%
\bibitem [{\citenamefont {Fiete}()}]{Fieteprivate}%
  \BibitemOpen
  \bibfield  {author} {\bibinfo {author} {\bibfnamefont {G.}~\bibnamefont
  {Fiete}},\ }\href@noop {} {}\bibinfo {note} {Private
  communication}\BibitemShut {NoStop}%
\end{thebibliography}
\end{document}